\documentclass[conference]{IEEEtran}
\usepackage{graphicx}
\ifCLASSINFOpdf
\else
\fi
\usepackage{amsmath, amssymb}
\usepackage{fixltx2e}
\usepackage{enumitem}
\usepackage{cite}
\usepackage{color}
\begin{document}
\setlength{\parskip}{0pt}

\title{Blindsight: Blinding EM Side-Channel Leakage Using Built-In Fully Integrated Inductive Voltage Regulator}

\author{\IEEEauthorblockN{Monodeep Kar\IEEEauthorrefmark{1},
Arvind Singh\IEEEauthorrefmark{1},
Santosh Ghosh\IEEEauthorrefmark{2},
Sanu Mathew\IEEEauthorrefmark{2},\\
Anand Rajan\IEEEauthorrefmark{2},
Vivek De\IEEEauthorrefmark{2},
Raheem Beyah\IEEEauthorrefmark{1} and
Saibal Mukhopadhyay\IEEEauthorrefmark{1}}
\IEEEauthorblockA{\IEEEauthorrefmark{1}School of Electrical and Computer Engineering, Georgia Institute of Technology, Atlanta, GA\\ 
Email: \{monodeepkar, rathorearvind19, rbeyah, saibal.mukhopadhyay\}@gatech.edu}
\IEEEauthorblockA{\IEEEauthorrefmark{2}Intel, Hillsboro, OR, Email: \{santosh.ghosh, sanu.k.mathew, anand.rajan, vivek.de\}@intel.com}}


%Telephone: (800) 555--1212, Fax: (888) 555--1212}


\maketitle

\begin{abstract}
Modern high-performance as well as power-constrained System-on-Chips (SoC) are increasingly using hardware accelerated encryption engines to secure computation, memory access, and communication operations. 
%High performance encryption is gaining sheer interest in the wake of protected video streaming, secure memory enclaves like SGX and high performance encryption of financial services. As the need for security grow tremendously in a data-driven world, securing the corresponding hardware platforms from a Side-Channel-Attack (SCA) will become crucial. Other than requirement for high-performance, another attribute of encryption system, which is critical in power constrained devices like IoTs, sensor nodes and wearables is high energy-efficiency. With increasing connectivity, the security of these systems are critical. 
The electromagnetic (EM) emission from a chip leaks information of the underlying logical operation being performed by the chip. As the EM information leakage can be collected using low-cost instruments and non-invasive measurements, EM based side-channel attacks (EMSCA) have emerged as a major threat to security of encryption engines in a SoC.  %Inductive IVR with randomized control loop has been shown to improve resistance against power attacks ~\cite{kar20178}. 
This paper presents the concept of \textit{Blindsight} where an high-frequency inductive voltage regulator integrated on the same chip with an encryption engine is used to increase resistance against EMSCA. High-frequency ($\sim$100MHz) inductive integrated voltage regulators (IVR) are present in modern microprocessors to improve energy-efficiency. We show that an IVR with a randomized control loop (R-IVR) can reduce EMSCA as the integrated inductance acts as a strong EM emitter and \textit{blinds} an adversary from EM emission of the encryption engine. The measurements are performed on a prototype circuit board with a test-chip containing two architectures of a 128-bit Advanced Encryption Standard (AES) engine powered by a high-frequency (125MHz) R-IVR with wirebond inductor. The EM measurements are performed under two attack scenarios, one, where an adversary gains complete physical access of the target device (EMSCA with Physical Access) and the other, where the adversary is only in proximity of the device (Proximity EMSCA). The resistance to EMSCA is characterized considering a naive adversary as well as a skilled one with intelligent post-processing capabilities. In both attack modes, for a naive adversary, EM emission from a baseline IVR (B-IVR, without control loop randomization) increases EMSCA resistance compared to a standalone AES engine. However, a skilled adversary with intelligent post-processing can observe information leakage in Test Vector Leakage Assessment (TVLA) test. Subsequently, we show that EM emission from the R-IVR \textit{blinds} the attacker and significantly reduces SCA vulnerability of the AES engine. A range of practical side-channel analysis including TVLA, Correlation Electromagnetic Analysis (CEMA), and a template based CEMA shows that R-IVR can reduce information leakage and prevent key extraction even against a skilled adversary.   

\end{abstract}


% \fi
\IEEEpeerreviewmaketitle

\section{Introduction}

\begin{figure}[!tb]
\setlength{\abovecaptionskip}{0pt}
\setlength{\belowcaptionskip}{0pt}
\centering
\includegraphics[width=3.3in,keepaspectratio]{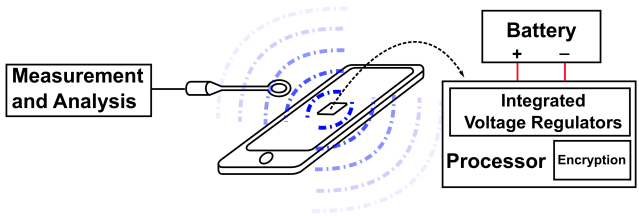}
\caption{EM attack on a Smart-phone}
\label{fig:1}
\end{figure}

High performance encryption is becoming a standard feature in modern hardware across different applications like protected video streaming ~\cite{pantos2017http}, data and memory protection ~\cite{rott2010intel, costan2016intel} and financial data transaction. One of the most common crypto algorithm to enhance security of servers, desktops, and mobile platforms \cite{rott2010intel} is Advanced Encryption Standards (AES). The latest processors and SoCs show a common trend of using dedicated accelerator for AES. A new instruction (AES-NI) has been added as an extension to x86 instruction set and is being widely used across Intel and AMD processors ~\cite{rott2010intel}. Similarly ARM cortex processor series also has dedicated instructions for AES and SHA-256.
% the  and modular Altera’s launch of OpenCL support for FPGA systems allows to achieve a throughput of 40 Gbits/second AES encryptions. 
The hardware acceleration of AES engines is also being actively studied for power constrained small IoTs edge devices to support secure communication, leading to design of energy-efficient AES engines ~\cite{mathew2015340,hamalainen2006design,moradi2011pushing}. As computing continue to become more ubiquitous, ensuring security of the encryption engines in computing and communication devices against side channel attacks (SCA) is becoming increasingly important and challenging. In particular, power dissipation and electromagnetic (EM) emissions from modern SoCs can leak compromising information and has emerged as key threats to security of modern SoCs. 
%As the need for security grow tremendously in a data-driven world, securing these hardwares from side channel attacks is an important aspect for maintaining integrity of these systems. 
%Although majority of the research has focused on characterizing and preventing information leakage in power consumption of the physical systems, the electromagnetic (EM) emission from these systems have also shown to be leaking compromising information. 
Consequently, power and EM side channel attack on AES architectures have received signifcant attention over last decade ~\cite{rott2010intel,longo2015soc,yamaguchi2010development,plos2008enhancing,ratanpal2004chip,wang2013role}. Although majority of the works focus on inhibiting power attacks, preventing EM attacks is gaining more importance in the era of mobile and ubiquituous computing, due to the simplicity and inexpensive nature of the EM attack. Mounting a power attack requires physical probing of the target device i.e. a direct access to the printed circuit board (PCB) and/or the exposed area of the package which houses the chip. EM side channel signatures on the other hand can be easily captured with in-expensive EM probes ~\cite{longo2015soc} (Figure \ref{fig:1}) by being in close proximity of the device.

Traditional approaches to EM countermeasures involve modifying the algorithm, architecture, or logic design of the AES engines. But the challenge comes from the significant power, performance, or area-overheads associated with these approaches, making adoption of these techniques unattractive to resource-constrained and performance-sensitive commercial products. Operating systems like Android5.0 and Apple IoS already suffer from slow memeory encryption, which suggests that additional performance penalty for side-channel security is unacceptable for commercial products ~\cite{gouvea2015implementing}. Another option of adding advanced EM shielding ~\cite{yamaguchi2010development,plos2008enhancing} comes at the expense of significantly increased packaging cost. Moreover recent EM attacks have been demonstrated on finished products with high-end packaging which came with EM shielding ~\cite{zajic2014experimental,do2013electromagnetic}. In essence, eliminating the physical leakage of EM signals is difficult and comes with high power and performance penalty and increased cost. Therefore, this paper pursues an orthogonal approach to thwarting EM side-channel attacks and develop {\it innovative techniques using existing components in modern SoCs to modulate information content in the EM signatures} and reduce information leakage.

A voltage regulator module (VRM) converts the input voltage from a voltage source (battery/power supply/harvested energy) to a suitable voltage for the application circuit. Traditionally VRMs are used as seperate integrated circuits (ICs) in the same board as the processor/SoC to generate different voltage levels. However, driven by the needs to (i) reduce noise in the power supply, (ii) enable fast dynamic voltage scaling for reducing power, and (iii) creating multiple voltage domain for efficient workload driven power management, there is a growing trend in integrating a VRM with the processors and SoCs in the same die ~\cite{kurd2015haswell,fluhr201512,bowhill2016xeon}. Inductive IVRs are switching voltage regulators with an inductance and a capacitance. The recent commercial processors like Intel Haswell and Xeon have demonstrated integration of inductive IVRs, on the processor chip~\cite{kurd2015haswell,krishnamurthy201720,krishnamurthy2014500,bowhill2016xeon}. In this paper, we argue that the EM signatures from a targeted platform are modified by the presence of an inductive IVR. The switching nature of operation and presence of an inductor, typically integrated close to the physical location of the application circuits creates an interference in the measured EM signatures. The interference is dictated by the current pattern through the inductor. 
%These IVRs enable a digital circuit to quickly and rapidly change its supply voltage depending on the throughput demand and enhance the energy efficiency of the overall system. Interestingly, from the context of EM emission, the presence of the inductor and the switching nature of the regulator emits strong EM radiations from processor which will interfere with the EM emissions due to logical operations (the {\it EM side-channel}). Moreover the small form factor of the inductor makes it difficult to capture relevant EM signatures from the processor activities.  

\noindent
\textbf{Motivation}: 

Kar et. al. in ~\cite{kar20178} presented power side-channel attack (PSCA) results for two configurations of an inductive IVR, namely a baseline IVR (B-IVR) representing typical operating mode of any inductive IVR and randomized IVR (R-IVR) where the control loop of the IVR is randomized. In general the techniques for improving PSCA resistance are not guaranteed to be effective for improving EMSCA resistance. However, the current transformations through an IVR, which are exploited for improving PSCA resistance, change the current pattern through the inductor and therefore carry the potential to be effective for improving EMSCA resistance as well. 

\noindent
\textbf{Contribution}: 

This paper presents the concept of \textit{Blindsight} where a high-frequency IVR is used to increase resistance against EMSCA. We, for the first time, experimentally characterize EM emission of a system-on-chip (SoC) with embedded encryption engine powered by an IVR. We demonstrate that an R-IVR reduces EMSCA as the integrated inductance acts as a strong EM emitter and \textit{blinds} an adversary from EM emission of the encryption engine. 

We consider two attack scenarios, namely, (i) EMSCA with physical access, where an adversary gained physical access to the device and performs localized EMSCA on the SoC using a probe with high spatial resolution; and (ii) Proximity EMSCA, where the adversary can only get to a close proximity of the target device and is forced to use a passive EM probe that can measure signature from a larger distance but with lower spatial resolution. We also consider adversaries with different skill-sets: 1) a naive adversary that can \textit{only} perform SCA on the raw EM signal captured by the probes (no post-processing skills) and 2) a skilled adversary who can perform intelligent post-processing on the captured data. 

Under the preceding attack model and adversary skills, we characterize EM leakage from a prototype board carrying a fabricated application-specific-integrated-circuit (ASIC) with two 128-bit AES engines, powered by B-IVR and R-IVR. We consider two architecturally different implementations of the AES-128 algorithm. The first design, suitable for a high throughput device such as desktop or server microprocessor is referred to as HP-AES, and the second design, suitable for a power constrained IoT device application is referred to as LP-AES. To quantify the EMSCA resistance, Correlation Electromagnetic Analysis (CEMA) which is a key-extraction attack and Test Vector Leakage Assessment (TVLA) which is a leakage analysis test are used. 

%
%attack  and evaluate the EM leakage from a prototype board carrying a fabricated ASIC, based on the circuit techniques proposed in ~\cite{kar20178}. The first scenario assumes that an adversary gained physical access to the device and performs EMSCA using a probe with high spatial resolution. The second scenario considers the situation where the adversary can only get to a close proximity of the target device and is forced to use a low resolution passive probe. For both these conditions, we evaluate the EMSCA resistance for the B-IVR and the R-IVR configurations. We also consider adversaries with different skill-sets: 1) a naive adversary with no post-processing skills and 
%b) a skilled adversary who can perform intelligent post-processing of data. To quantify the EMSCA resistance, Correlation Electromagnetic Analysis (CEMA) which is a key-extraction attack and Test Vector Leakage Assessment (TVLA) which is a leakage analysis test are used. 
% 
%We chose two architecturally different implementations of the AES-128 algorithm with different inherent EMSCA resistance, to demonstrate the results. The first design, suitable for a high throughput application is referred to as HP-AES and the second design, suitable for a power constrained application is referred to as LP-AES. 

The measurements and analysis performed in the paper demonstrate following key observations:  
\begin{itemize}[leftmargin=*,noitemsep,topsep=0pt]
\item{If IVR is not used (standalone AES mode), a naive (and skilled) adversary can extract useful information from EM leakage both with physical access, and from close proximity. The measurement shows that using CEMA, the secret keys from HP-AES and LP-AES engines can be extracted from 40,000 and 1,000 traces. As expected, the TVLA shows very strong information leakage. } %%% SM - Do you mean CPA or CEMA

\item{If the attacker has \textit{physical access of the device and high resolution spatial probe} a skilled adversary can measure EM signatures by placing the probe near specific pins of the SoC package. In particular, we show that placing the probe near the inductor node and the supply node of the IVR shows strong information leakage in the B-IVR mode in a TVLA test. However, no information leakage is measured at the same locations when R-IVR is used.} %%% SM- Include the CPA number of traces here

\item{If the attacker can only come in the \textit{proximity of the device and hence, uses low-resolution probe}, a naive adversary can extract secret key in the standalone mode, but fails to extract key with IVR. A skilled adversary with appropriate post-processing, can extract information from the EM signatures in B-IVR mode in TVLA test for both HP-AES and LP-AES. Moreover a successful CEMA was observed on LP-AES with 40,000 traces, although HP-AES failed to show a successful CEMA with 500,000 traces. However, even for a skilled adversary, the R-IVR mode suppresses information leakage and TVLA shows no noticeable leakage with linear and higher-order statistics and CEMA was not successful even with 500,000 traces, for both AES designs. 
We proposed a new attack model by subtracting a template EM signature from the measured signatures to remove the effect of randomization, but no successful CEMA attack was observed.}
\end{itemize}

%In particular, the major contribution of this paper are the following:
%\begin{itemize}
%\item{The paper presents an in-depth measurement-based analysis to evaluate potential of inductive IVRs in reducing EM information leakage. We fabricate a test-chip in 130nm containing two 128-bit AES engines supplied by a IVR with package bondwires as inductance and characterize EM signatures from the entire system.} 
%\item{We use EM probes with different resolution to characterize leakage from different physical location of the prototype system. We use Correlation Power Analysis (CPA) which is a key-extraction attack and Test Vector Leakage Assessment (TVLA) which is a leakage analysis test, to evaluation of EM leakage. The captured EM signatures are analyzed across different frequency bands upto 500MHz.}
%\item{The EM leakage behavior of different AES architectures are found to be drastically different. Before turning on the randomized mode of the IVR, TVLA results show that the system EM signatures for both AES designs shows signs of leakage. However CPA results show that the HP-AES is more robust compared to LP-AES. In the randomized mode, BlindSide successfully prevents TVLA leakage and CPA for both AES designs, even under advanced attack methods like template attack.}
%
%\end{itemize}
%
In summary, we show that, in addition to the performance efficiency gained by the use of an IVR, various naive side-channel attacks performed by measuring EM signature from close proximity can be thwarted. A skilled adversary can still extract information when a baseline IVR architecture is used. However the R-IVR reduces vulnerability to EMSCA, even for a highly-skilled adversary, by using a minor design modification on the existing IVRs. 

The rest of the paper is organized as follows: Section \ref{sec:background} provides background on side-channel attack; Section \ref{sec:IVR_SCA} discusses the preliminary concepts on the role of IVR in power side-channel attack; Section \ref{protosystem} describes the design of the prototype system; Section \ref{sec:em-characterization-forensic}, section \ref{sec:results} and section \ref{sec:proximity} present the measurement results  corresponding to the two attack scenarios described in this paper; Section \ref{sec:discussion} discusses additional topics on the proposed method and Section \ref{conclusion} concludes the paper.

%% SM - I dont see any conclusion

\section{THREAT MODEL AND PRELIMINARIES}
\label{sec:background}

\subsection{Threat Model: EMSCA}
\begin{figure}[!tb]
\setlength{\abovecaptionskip}{0pt}
\setlength{\belowcaptionskip}{0pt}
\includegraphics[width=3.3in,keepaspectratio]{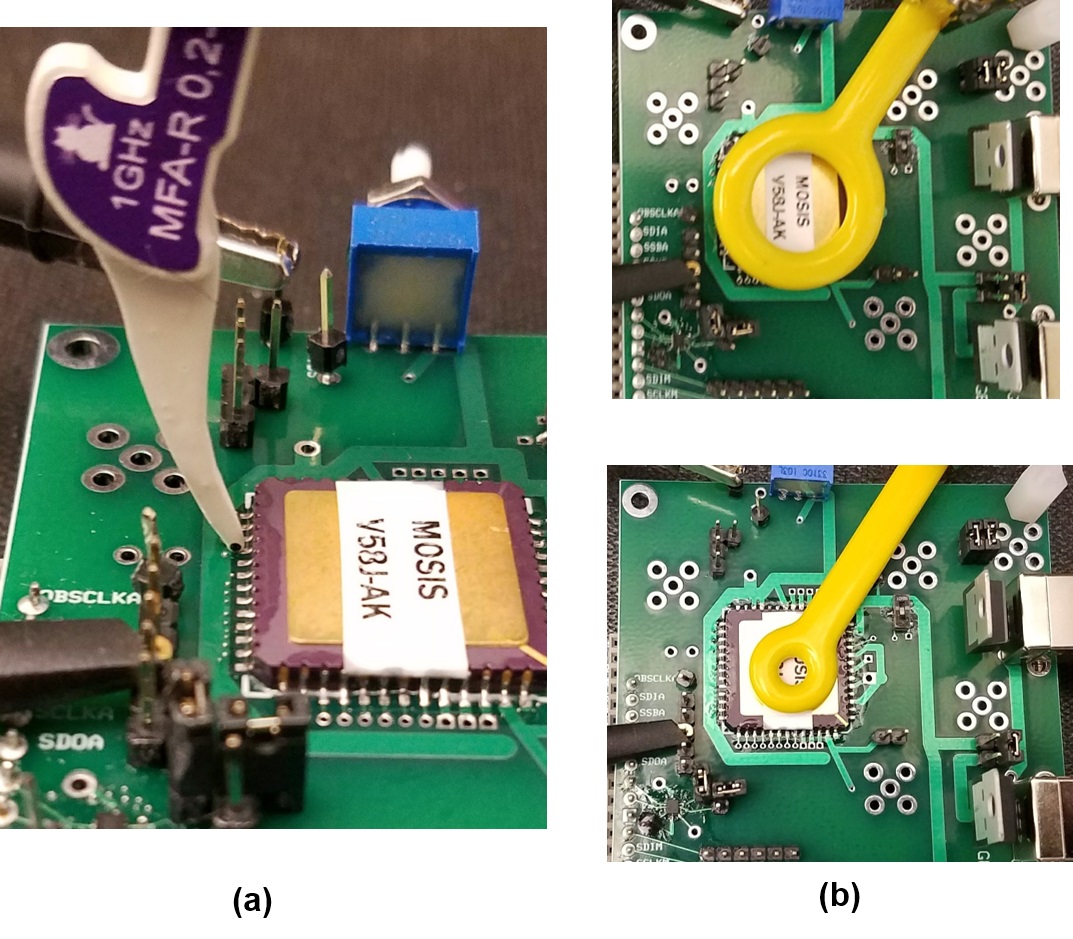}
\caption{Different EM probes used for analysis (a) A active probe with high spatial resolution for EMSCA with physical access(b) Passive probes for proximity EMSCA}
\label{fig:Probe_Pics_Combined}
\end{figure}
Due to the abundance of the connected devices as well as their expected hostile operating conditions without any supervision, it is becoming increasingly easy to snoop side channel signatures from a device. The most exploited side channels are power i.e. current flowing into the supply and ground pins of the targeted hardware and EM emissions from the targeted system during the encryption process. For collecting power traces, it is necessary for the adversary to make physical contact at the power pin of the target platform and is difficult to perform on a finished product. Picking up EM signatures from a device is non-invasive in nature and therefore fits perfectly for a run-time attack on a device. The past research shows successful key-recovery for both symmetric ciphers (AES) and asymmetric ciphers (RSA, ECC etc.) ~\cite{korak2012attacking,genkin2016ecdh,genkin2015stealing}. 

Several tiers of adversaries and attack scenarios have varying access/proximity to the device to be attacked. Depending on the proximity to the devices, an intelligent adversary will also select an appropriate EM probe for the attack. We envision two attack scenarios and choose appropriate probes for each of them. 

\begin{itemize}[leftmargin=*,noitemsep,topsep=0pt]
\item{\textbf{EMSCA with Physical Access}: In the first scenario, we envision that the device has been captured by the adversary and the adversary has the ability to deconstruct the device and have direct access to the pins and traces. We have used a Langer MFA-R near field probe with a 300$\mu$m resolution and an active low noise amplifier to characterize the EM leakage from different pins of the package as shown in Figure \ref{fig:Probe_Pics_Combined}a. The probe has a bandwidth of 6GHz which allows accurate measurement of the high frequency EM radiations. As pointed out in ~\cite{balasch2015dpa}, EM probes like this indirectly measure the power signature from the corresponding pin or PCB trace. Due to high sensitivity to the distance between the probe and the package pin, the probe has to be placed right on top of the pin, as shown in Figure \ref{fig:Probe_Pics_Combined}a.} 

\item{\textbf{Proximity EMSCA}: The second attack scenario considers a case when the adversary does not have access to the actual device, but can come within close proximity of the device to be attacked. For example, the adversary can be standing in line behind the victim, having an actual conversation with the victim, or can place an inconspicuous item containing a probe (as done in ~\cite{genkin2015stealing} by hiding a EM probe within a Pita bread). Figure \ref{fig:Probe_Pics_Combined}b shows two passive EM probes by Beehive Corp. with significantly large loop area than the Langer probe described earlier. The loop diameters are 0.85 in and 0.4 in for the larger and the smaller loop respectively. The probe output powers into a 50 ohm load. These inexpensive probes are easy to acquire and hide. Both the probes are placed on top of the package at different locations for characterizing EM leakage and will be described in further detail in section \ref{probe-characterization}.} 

\end{itemize}

\subsection{State of the Art: EM Countermeasures}

%\subsubsection{CPA Using Templates}
%In a superimposed EM signature, patterns contributed by the inductor emission is repetitive. In the R-IVR mode, the randomization is introduced by a pseudo-random source which repeats the pseudo-random sequence after a certain interval. A template attack performs well for cases where the repeated measurement of a single event produces different traces. An average template is constructed by aligning and averaging the EM signature for multiple traces over a window length equal to the period of the pseudo-random source. The period can be determined by visual inspection of the captured traces or through frequency domain representation of the data. The average template consists of two parts: 1) the pseudo-random EM signature in absence of any AES operation and 2) an averaged EM leakage from the AES operations. We note that this is not simple addition of these two components as the IVR control loop reacts to the current variation due to different AES encryptions as well as the output of the pseudo-random sequence. The averaged template is then subtracted from the original traces and a CPA is performed on the resultant data. 

\begin{figure}[!tb]
\setlength{\abovecaptionskip}{0pt}
\setlength{\belowcaptionskip}{0pt}
\includegraphics[width=3.3in,keepaspectratio]{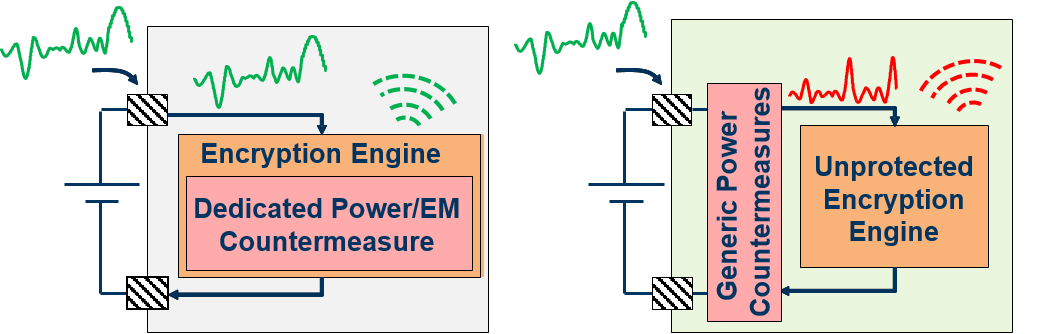}
\caption{Effect of different countermeasures for PSCA on EMSCA}
\label{fig:power_EM_CM}
\end{figure}

Countermeasures are modification in the design of the hardware to reduce side channel vulnerability.  Different countermeasures have been proposed by the researchers in past decade to prevent side channel leakage, both for PSCA and EMSCA. Majority of these countermeasures target power attacks and aim at decorrelating the measured power signatures and data at the intermediate steps of the algorithm. This can be achieved by changing the intermediate steps of the encryption algorithm, changing the architecture or using logic styles where the power consumption is unrelated to the switching activity. Each of these techniques change the design of the hardware either in algorithm, architecture or physical implementation level. A parallel category of PSCA countermeasures does not modify the design of the encryption engine, rather uses generic techniques like attenuation, noise addition and transformations for reducing the correlation. 

The nature of power and EM side channel are radically different, therefore countermeasures for PSCA might not be effective for EMSCA and vice-versa. In general, any PSCA countermeasure that depends on isolation or attenuation of the power signatures ~\cite{tokunaga2010securing,wang2013role,yu2015leveraging} and does not modify the design of the encryption engine, may not reduce EM-based side channel leakage as explained in Figure \ref{fig:power_EM_CM}. This is due to the fact that the leakage is not eliminated at source and an EMSCA adversary has the location of the probes as another degree of freedom. Therefore the adversary can capture signature from a physical location which bypasses the effect of many of these techniques. 

For reducing EM leakage, one simple yet elegant solution is to use any form of shield on top of the targeted device as proposed by  Plos et. al. in ~\cite{plos2008enhancing}, however the solutions are ad-hoc and difficult to achieve for a mass scale commercial production. Poucheret et. al. proposed distribution of the leaking electrical paths throughout the physical implementation of the hardware to prevent EMSCA ~\cite{poucheret2010spatial}. Doulcier-Verdier et. al. used duplicated-complimented logic style to prevent both PSCA and EMSCA ~\cite{doulcier2011side}. A serious bottleneck of these types of countermeasures is the energy-efficiency and design complexity of the proposed techniques. While encryption bit-rate is critical for high-performance systems, low-power devices require lower-energy per encryption. Most of the proposed countermeasures suffer from a performance penalty due to added complexity for side channel protection. Moreover the design and validation effort needed for incorporating these countermeasure techniques further make them unattractive for use in general purpose products. Therefore there is a critical need for finding a unique low cost solution for addressing both PSCA and EMSCA.

\subsection{Power Delivery and Voltage Regulators}

\begin{figure}[!tb]
\setlength{\abovecaptionskip}{0pt}
\setlength{\belowcaptionskip}{0pt}
\includegraphics[width=3.3in]{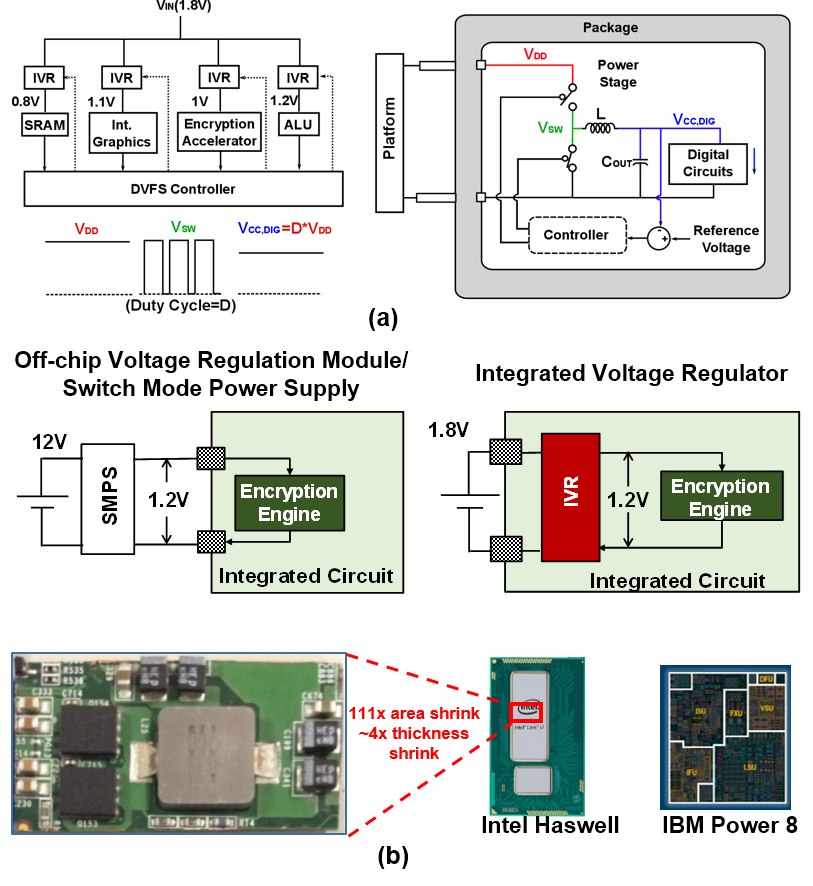}
\caption{(a) Circuit and system level diagram of inductive voltage regulators (b) Traditional off-chip power delivery architecture vs. integrated voltage regulators }
\label{fig:IVR_Intro}
\end{figure}

Processors and SoCs require multiple supply voltages, also known as power rails, for optimizing energy efficiency across different operating conditions. Voltage regulators are therefore one of the key components in the power delivery architecture of a processor. Inductive voltage regulators are a popular class of switching voltage regulators and widely used for their superior power efficiency compared to other classes of VRM.

\noindent
\textbf{Working Principle of Inductive Regulators}
Figure \ref{fig:IVR_Intro}a shows the circuit diagram of an inductive regulator. The switches M\textsubscript{1} and M\textsubscript{2} are continuously driven by two square waves at frequency F\textsubscript{SW}. The duty cycle of the square waves determines the output voltage. The inductor (L) and the capacitor (C\textsubscript{OUT}) create a bandpass filter whose cutoff frequency (F\textsubscript{LC}) is lower than the switching frequency (F\textsubscript{SW}). The switching node V\textsubscript{SW} resembles a square wave which is filtered out to create a steady DC voltage V\textsubscript{OUT}. The output node drives different digital blocks of a microprocessor or SoC. Every voltage regulator requires a controller which ensures that if the load current demand increases, the regulator can supply the required current without the output voltage dropping. A feedback controller senses the difference between the output voltage and the reference voltage and adjusts the duty cycle to set the output voltage at the desired value.

\begin{figure}[!tb]
\setlength{\abovecaptionskip}{0pt}
\setlength{\belowcaptionskip}{0pt}
\includegraphics[width=3.3in]{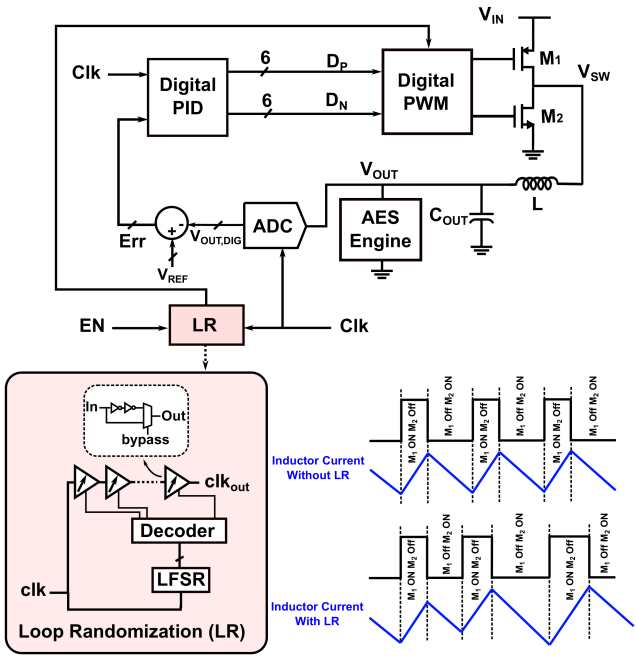}
\caption{Architecture of a security aware inductive IVR ~\cite{kar20178}. Loop randomizer circuit and its effect on the inductor current are shown.}
\label{fig:ivr_arch_details}
\end{figure}

\noindent
\textbf{Integrated Inductive Voltage Regulators}
\label{ivr}
Traditional power delivery architecture consists of multiple voltage regulator ICs, typically present in the motherboard/logic board. Integrated voltage regulators (IVR) are voltage regulators integrated with the digital circuits in the same silicon die. IVRs reduce the volume and complexity of power delivery for multiple supply voltages. Among various popular topologies of IVR, on-chip low dropout regulators (LDO) have been used across multiple generations of processors and SoCs. However LDOs have poor power efficiency compared to switching regulators (inductive buck and switched-capacitor) across a wide range of voltage-frequency (V-F) states. Innovations in integrating tiny passives (inductance and/or capacitance) with digital transistors in the same silicon die or same package ~\cite{lambert2014package}, enabled usage of inductive IVRs in commercials products ~\cite{kurd2015haswell}. The reduced value of the inductance requires the IVRs to switch at very high frequency ($\sim$100 of MHz), much higher than conventional off-chip VRMs ($\leq$1MHz). The switching frequency of IVRs is closer to the operating frequency of digital circuits.

\section{IVR AND EM SIDE CHANNEL ATTACK}
\label{sec:IVR_SCA}
This section motivates the potential use of inductive IVRs to inhibit EM based side-channel attack. To motivate IVR based EMSCA robustness, the section first summarizes the impact of IVR on PSCA, and a security-aware IVR architecture to reduce power side-channel leakage, as presented in ~\cite{kar20178}. Subsequently,  the impact of an inductive IVR on EM side channel leakage is elaborated and it is shown that the IVR properties which improves PSCA robustness can also be effective for improving EMSCA robustness.

\subsection{IVR and Power Side-Channel Attack}
\label{subsec:IVR_PSCA}
%\noindent
%\textbf{Why IVRs can protect against PSCA}
Kar et. al. in ~\cite{kar20178} demonstrated that presence of an inductive IVR affects the power side-channel leakage of a platform. The authors made the observation that when an inductive IVR supplies power to an encryption engine, the current signature at the IVR output is isolated from the IVR input i.e. the current drawn from the supply of the regulator (battery for a laptop/handheld device and mains supply for a desktop). However the input current is not completely independent of the load current, rather it is a transformed version of the load current. The improvement in the PSCA resistance is governed by three different transformation of IVR's load current to the input current.
\begin{itemize}[leftmargin=*,noitemsep,topsep=0pt]
\item{Large Signal Transformation}: The continuous switching of switches M\textsubscript{1} and M\textsubscript{2} in the power stage creates a switching current pattern at the IVR input.\\
\item{Small Signal Transformation}: The load current signatures are filtered by the frequency dependent transfer function of the PID compensator in the feedback loop.\\
\item{Misalignment}:The IVR switching clock is asynchronous w.r.t the clock driving the encryption engine. The asynchronous nature of these two clocks causes a one to many mapping from load current to input current.
\end{itemize}

\subsection{A Side-Channel-Security-Aware IVR}
\label{subsec:secure_IVR}
\label{sec-ivr}
Figure \ref{fig:ivr_arch_details} shows the overall architecture of a side-channel-security aware IVR architecture, as presented in ~\cite{kar20178}. The power stage of the IVR switches at 125MHz switching frequency and package-bondwires are used as inductance. The capacitor of the power stage is embedded within the die. The IVR uses a digital feedback loop as a controller. A digital controller first digitizes the output voltage using a high-speed analog-to-digital converter (ADC) and the control algorithm (proportional-integral-derivative (PID)) is implemented in digital logic. The controller output is fed to a block called digital pulse width modulator (DPWM) which converts the digital input to the duty cycle of the square wave. All these aforementioned blocks are part of any typical IVR architecture and therefore it is referred to as a Baseline IVR (B-IVR) mode. 

The design also contains an extra circuit called loop-randomizer (LR), as described in ~\cite{kar20178}. LR inserts delay into IVR's control loop by delaying the power stage clock through a chain of delay-elements. Each delay element can be set in a bypass mode where the input signal bypasses the inverter. A 4-bit maximal length LFSR generates a sequence of 15 pseudo-random outputs which determines how many inverters are bypassed in the entire chain. LR creates a pseudo-random perturbation in the IVR's output voltage as well as its inductor current, as shown in Figure \ref{fig:ivr_arch_details}. The mode when LR is enabled is referred to as a randomized-IVR (R-IVR). 

\begin{figure}[!h]
\setlength{\abovecaptionskip}{0pt}
\setlength{\belowcaptionskip}{0pt}
\centering
\includegraphics[width=2.5in]{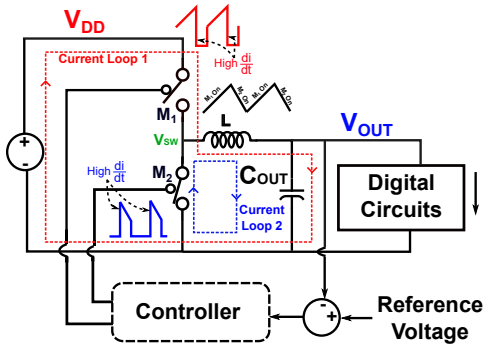}
\caption{Current loops in an inductive regulator generating EM interference}
\label{fig:ivr_em}
\end{figure}
\subsection{IVR and EM Side-Channel Leakage}
\label{subsec:IVR_EMSCA}
EM radiation can be generated by two sources: Alternating electric field source (high impedance) or alternating magnetic field source (low impedance). An inductive regulator has two main loops where high AC currents flow as shown in figure \ref{fig:ivr_em}. When the high-side switch M\textsubscript{1} is on, the current flows from supply via M\textsubscript{1} and L to the C\textsubscript{OUT} and the load. The current flows back via ground to the input. The AC portion of the current will flow via the input and output capacitors (Figure \ref{fig:ivr_em}). When M\textsubscript{1} switches off, the inductor current will keep flowing in the same direction, and the low side switch M\textsubscript{2} is switched on. The current flows via M\textsubscript{2}, L to the C\textsubscript{OUT} and the load and back via ground to M\textsubscript{2}. This loop is shown in blue. Both these loops carry discontinuous currents, meaning that they have sharp rising and falling edges at the beginning and end of the active time. These sharp edges have fast rise and fall times (high di/dt). Therefore they have a lot of high frequency content.

Keeping EMSCA in context, these properties of an inductive IVR make it unique compared to a LDO or a switched-capacitor regulator. In-fact meeting the electromagnetic compliance, as guide-lined by Federal Communication Commission (FCC), of inductive regulators is a major design challenge. However for EMSCA protection, the same interference can be exploited to the designers' advantage.

\begin{figure*}[!h]
\setlength{\abovecaptionskip}{0pt}
\setlength{\belowcaptionskip}{0pt}
\includegraphics[width=\textwidth]{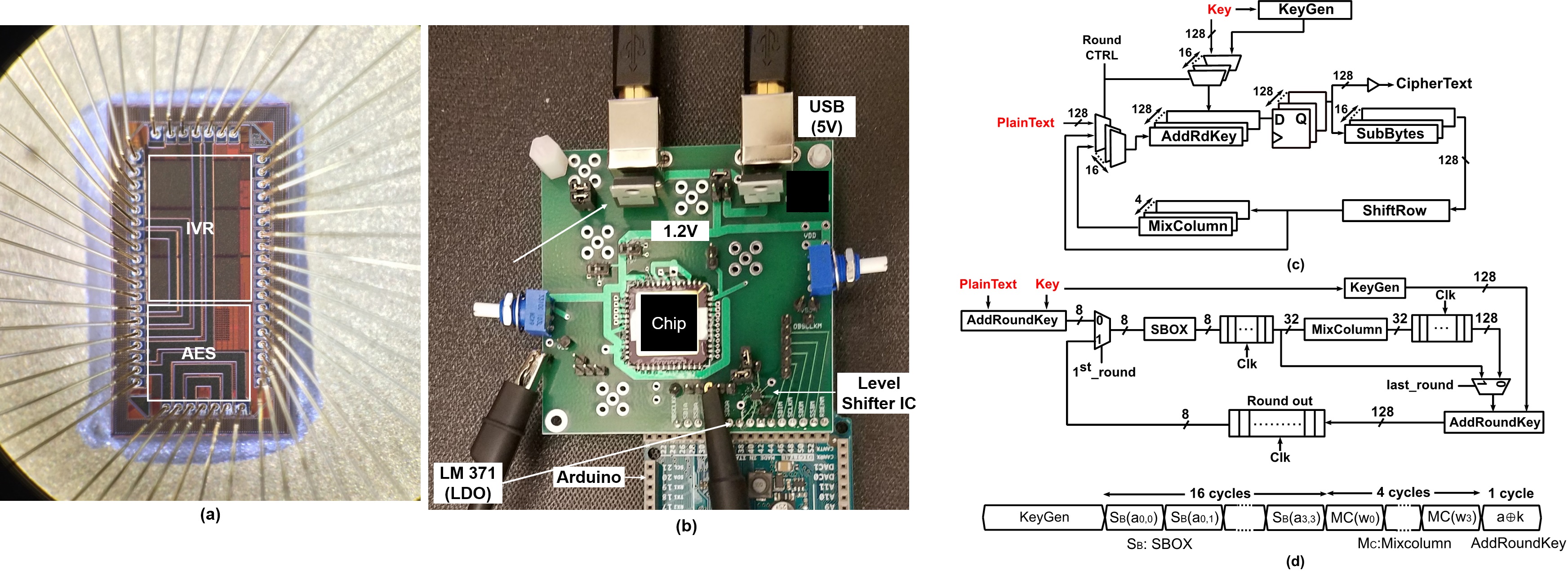}
\caption{(a) ASIC Micrograph with bondwires (b) Prototype PCB for characterization (c,d) Architectures of the implemented AES engines}
\label{fig:4}
\end{figure*}

\subsection{Motivation and Contribution}
\label{subsec:problem}

The EM emission from the inductor is not guaranteed to improve the EMSCA resistance, if the probing location can be adjusted to pick up the signatures from the AES engine without any interference from the inductor. This can happen if the inductor is physically distant from the electrical paths of the AES engine, and is true for commercial processors where the voltage regulator IC is on the same board, but physically distant from the processor. However, in any integrated VR, as in recent processors such as Haswell, the small form factor of the inductor ensures a compact placement close to the load circuit. Therefore it is difficult to separate out the effect of the inductor from that of the AES engine in the captured EM signatures. As the interference from inductor is a direct function of the current flowing through it, any properties of the inductor current are critical to analyze the EMSCA resistance of such a system. Moreover, the IVRs operate at frequencies ($\geq$100MHz) much closer to the processor's clock frequency (~1GHz), compared to off-chip VRs (~1KHz). Hence, the EM emissions from inductors in IVR are likely to more strongly interfere with EM emissions from the processors. Therefore, although off-chip VRs have shown to have little effect in reducing information leakage from the processor (in fact, in certain cases, off-chip VRs have shown to be a major source of leakage ~\cite{saab2016key}), the same conclusion cannot be drawn for on-chip VRs. This paper for the first time presents an in-depth measurement and characterization of the effect of IVR on EM leakage from SoCs. 

Although the architecture presented in ~\cite{kar20178} is focused on protection against PSCA, we observe that the inductor current of an IVR is also a function of the IVR input current. As the EM emission from the inductor is also a direct function of the current flowing through it, the architecture presented in ~\cite{kar20178} is relevant to EMSCA as well. However the effect on EMSCA resistance of the system is not addressed by the authors in ~\cite{kar20178}. In this paper, we aim to perform EMSCA analysis on a system of an inductive IVR and an AES engine for the B-IVR and the R-IVR modes. The experimental characterization is performed on a prototype circuit board composed of an ASIC, fabricated in 130nm, containing two architectures of AES-128 algorithm and the inductive IVR architecture proposed in ~\cite{kar20178}. The prototype system represents a microcosm of a high-performance or low-power SoC with hardware acceleration for AES encryption. More importantly, the prototype makes the AES engines more vulnerable as it does not include noise from other components in a chip.

%% SM - Use the Section THreat Model Here
%In this section introduce the EM attacks, different types of probes [just show their picture not the positions as in 7(left) and 8(top). Then explain the attack scenario, mention the raw attack will be on the measured signature, explain signal post-processing, explain the statistical tests including definition of template attack. 

\section{PROTOTYPE SYSTEM}
\label{protosystem}

\subsection{System Design}
\label{sysdesign}

Figure \ref{fig:4} shows the prototype board for evaluation. The designed ASIC is powered by standard USB connections. An off-chip voltage regulator (LM317) is used to convert 5.0V from the USB to 1.2V supply for the ASIC. The off-chip voltage regulator represents a traditional off-chip power delivery architecture. Even if IVR is present in a processor/SoC, an off-chip VRM is still needed to convert the platform input voltage to a tolerable input voltage of the IVR. The plaintexts and key of AES encryptions are written within the ASIC using an Arduino through a standard serial-to-parallel-interface (SPI).

\subsection{Architecture of the ASIC}
\label{asic-arch}

The ASIC has two architectures of the AES-128 algorithm. The die photo of the ASIC is shown in Figure \ref{fig:4}. LR is run at 1/8th of the IVR's sampling frequency. Therefore, the control loop delay changes once every 4th switching cycle of the IVR. LR creates a pseudo-random perturbation in the IVR's output voltage as well as its inductor current.

The AES-128 algorithm has a 10-round operation. The first architecture, referred to as high-performance AES (HP-AES), executes each AES round (all 16 bytes of the intermediate state) in one cycle (Figure \ref{fig:4}c) ~\cite{satoh2001compact}. The latency for one encryption is 11 cycles which makes the HP-AES suitable for latency-critical applications like memory encryption. The second AES architecture is referred to as low-power AES (LP-AES) as it is suited more for a light-weight low-power application (Figure \ref{fig:4}d). The datapath consists of a single S-BOX, 128 XORs for AddRoundKey, a word mix-column unit and intermediate registers for data storage. The bytes of the intermediate states are processed serially, causing a higher latency per encryption. The silicon area is significantly lower which makes the LP-AES architecture suited more for edge devices like wearables and sensor nodes. However designs similar to LP-AES, where rounds are executed serially, are found to be more vulnerable to correlation based attacks ~\cite{mukhopadhyay2014hardware,singh2015exploring}. The round-keys for both architectures are generated on the fly. 

\subsection{Packaging}
\label{pkg}
Each silicon die is accompanied with a package which forms the connections with the PCB. Packages play a critical role in leakage of EM side channel signature as different components of the package, mostly the parasitic inductance can amplify or mask the desired signatures. The ASIC is packaged in a Leadless Ceramic Package (LCC). The pads in the die are attached with the package with bondwires. Each bondwire is 5.5mm long, 1.3mil thick and offers roughly 5.8nH inductance. As the package is leadless, minimal inductance is offered by the connections between the PCB and the package.

\begin{figure}[!tb]
\setlength{\abovecaptionskip}{0pt}
\setlength{\belowcaptionskip}{0pt}
\includegraphics[width=3.3in,keepaspectratio]{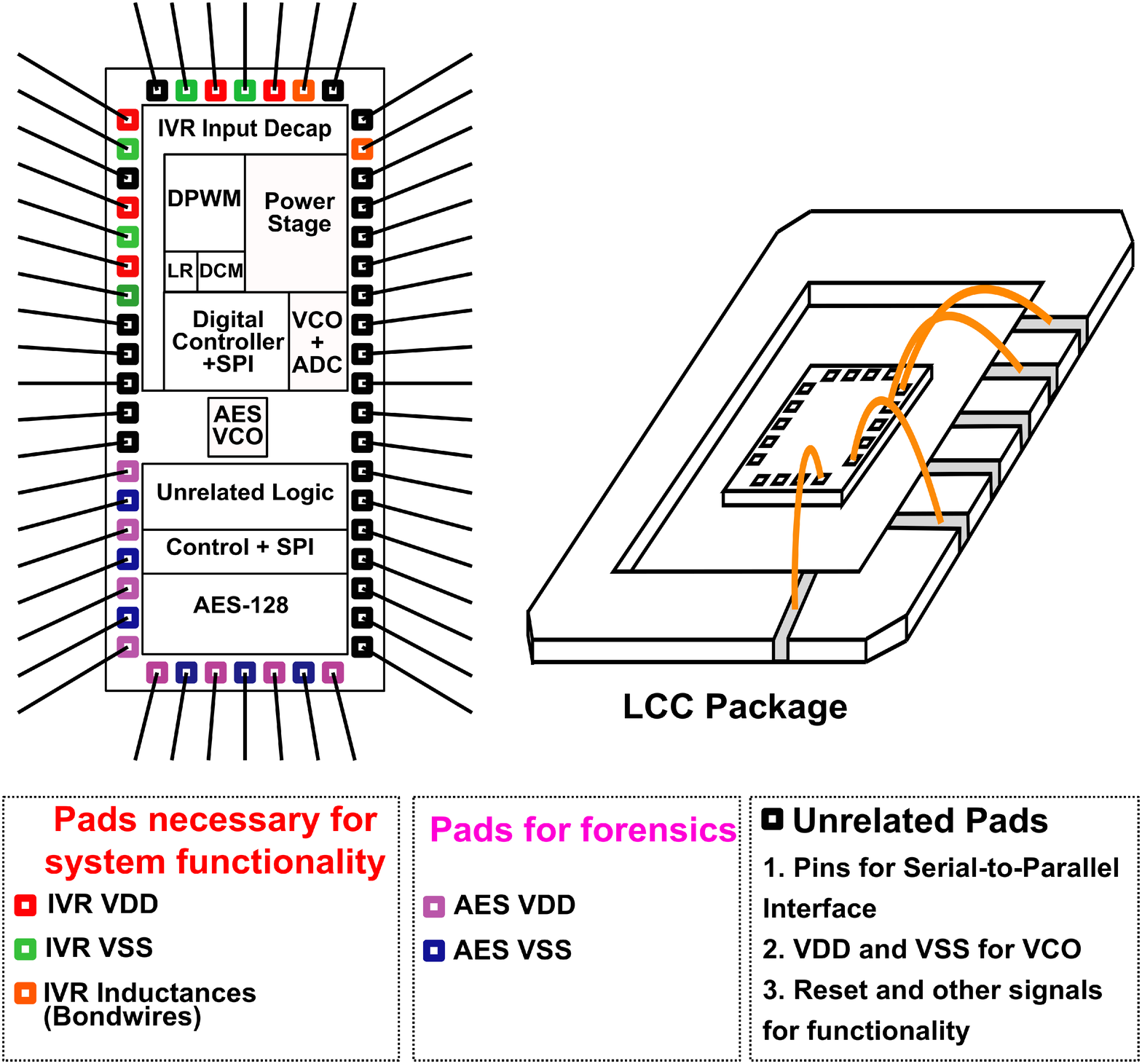}
\caption{Pad assignment of the fabricated ASIC and the corresponding LCC package}
\label{fig:5}
\end{figure}

In any general purpose hardware platform, the details of the pin mapping of the processor/microcontroller and the PCB traces are publicly released. When such a system is attacked, these information are typically exploited by the adversary to find out the suitable points for probing. For example, authors in ~\cite{balasch2015dpa} use the decoupling capacitor close to the microcontroller core to pick up the EM signatures. We assume that the pin mapping and the PCB routing of the prototype system is known to an adversary. Figure \ref{fig:5} shows the pads of the ASIC and their corresponding pins in the package. The IVR input, ground and the indcutor pins are towards the top-right corner of the chip. In order to characterize the AES without the effect of the IVR, the power (V\textsubscript{DD,AES}) and ground (V\textsubscript{SS,AES}) pin of the AES are separately connected to the package. These pins won't be present in a commercial chip: the power pin would effectively be the IVR output and the ground pin would be shorted internally to the IVR ground. The pins which do not carry side channel signatures are marked in black. The parasitic inductance and resistance of a LCC package are significantly higher than the advanced packages like flipchip/C4. Although using this package enabled us to exploit the package bondwires as IVR inductance, the higher inductance of the bondwires connecting AES supply and ground to the package creates EM emission directly from the AES engine, even when the IVR is supplying the AES. Therefore, enhancing EM side channel resistance for this prototype is more challenging compared to a commercial IVR which would use some form of integrate inductance (spiral inductance, silicon interposer, on-die solenoid) in an advanced package.

\subsection{Measurement Cases}
\label{cases}

\begin{figure}[!tb]
\setlength{\abovecaptionskip}{0pt}
\setlength{\belowcaptionskip}{0pt}
\includegraphics[width=3.3in,keepaspectratio]{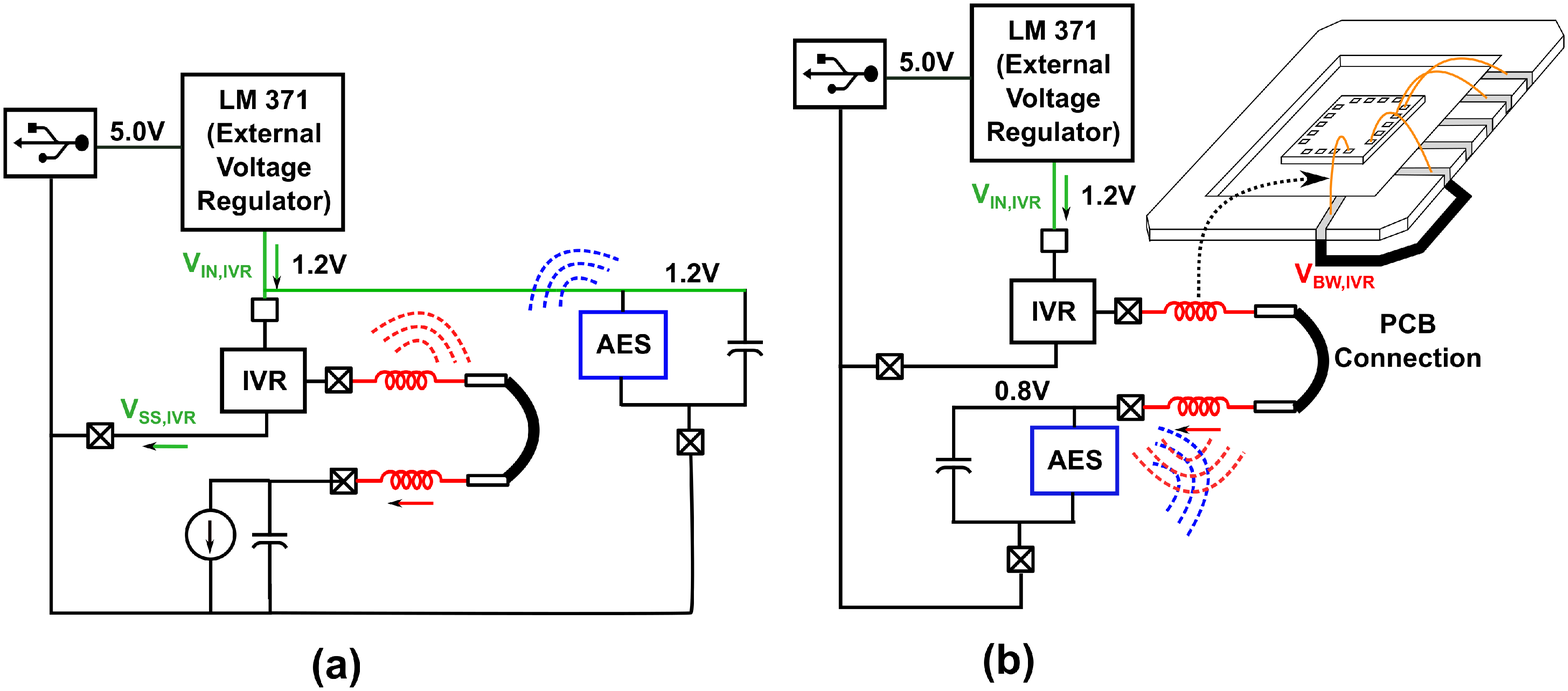}
\caption{Measurement scenarios a) AES is powered by an external voltage-regulator and b) IVR is powering AES engine}
\label{fig:6}
\end{figure}

Measurements were carried out at two different scenarios as depicted in Figure \ref{fig:6}. 
\begin{itemize}
\item{\textbf{Standalone AES}: The AES block and other peripheral digital circuits are powered by the off-chip voltage regulator (LM317). This mimics a traditional power delivery architecture. To prove the point that having a strong EM radiator near the encryption engine will have insignificant effect on the EM leakage, we keep the IVR on i.e. the IVR drives a steady load current. The switches M\textsubscript{1} and M\textsubscript{2} switch continuously. Naturally the inductor carries switching current and radiates strong EM signatures. However, as the IVR does not supply the AES engine, the inductor current and the corresponding EM emission have no relation to the AES current.}
\item{\textbf{IVR-AES}: The AES block is powered by the IVR. In this mode, the emission from the IVR inductor is linked with the AES activities. We evaluate the following two modes for IVR-AES: in B-IVR, the LR is disabled and in R-IVR, the LR is enabled which randomizes the control loop.} 
\end{itemize}  

\subsection{Measurement Details}
\label{probes}
\begin{figure}[!tb]
\setlength{\abovecaptionskip}{0pt}
\setlength{\belowcaptionskip}{0pt}
\includegraphics[width=3.3in,keepaspectratio]{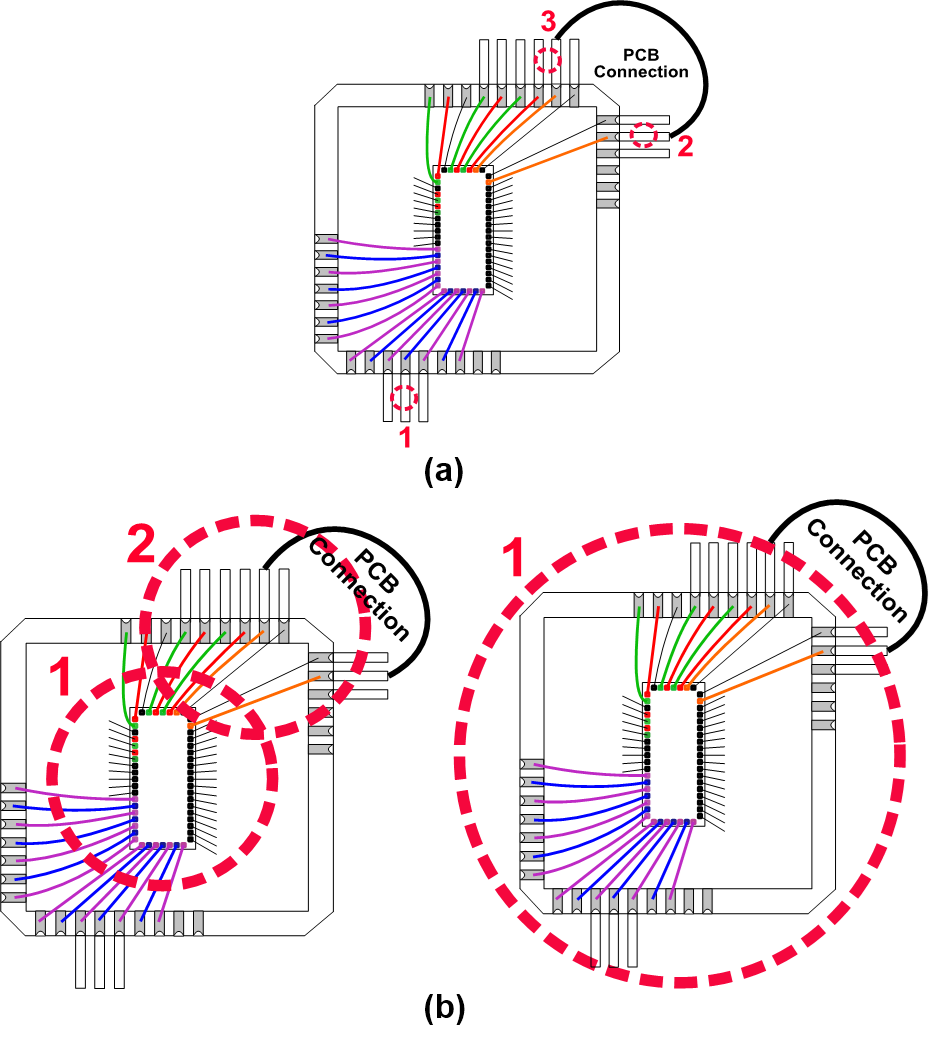}
\caption{Probing locations for different attack scenarios (a) EMSCA with Physical access (b) Proximity EMSCA}
\label{fig:Probe_locations_compiled}
\end{figure}

{\bf Placements of probes}: Figure \ref{fig:Probe_locations_compiled}a shows the placement of the high resolution active probe near the pins of the prototype test-chip. Figure \ref{fig:Probe_locations_compiled}b shows the potential placement options for the passive probes. The large loop probe spans the entire package and hence, has one placement location (location 1) as shown in figure \ref{fig:Probe_locations_compiled}b. The small loop probe has a higher bandwidth and provides more resolution in the placement of the probe (location 1 and 2, in Figure \ref{fig:Probe_locations_compiled}b).

\begin{figure*}[!tbhp]
\setlength{\abovecaptionskip}{0pt}
\setlength{\belowcaptionskip}{0pt}
\centering
\includegraphics[width=5.8in]{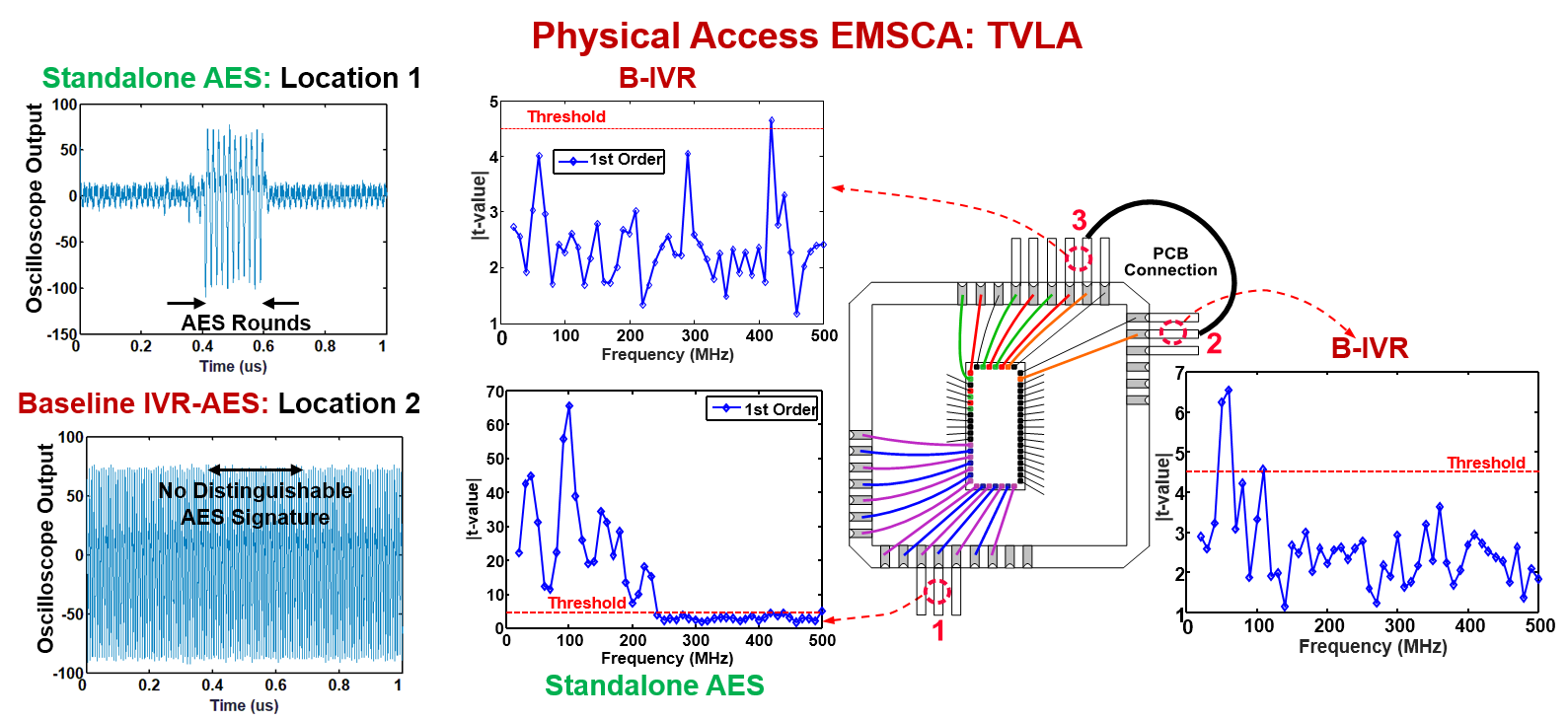}
\caption{(a) Sample traces captured in EMSCA with physical access from different pins of the chip and (b) the corresponding TVLA results}
\label{fig:forensics_results_compiled}
\end{figure*}

{\bf Statistical Tests}: The commonly used SCA resistance quantification approaches focus on an adversary's ability to extract the unknown key of an encryption engine. Both CEMA and differential elctromagnetic  analysis (DEMA) have been used as key extraction attack. Le et al. ~\cite{le2006proposition} have shown correlation based attack to be more efficient than a DEMA or DPA approach. A CEMA uses Pearson’s correlation between the measured side channel traces and a power-model to extract the secret key. The power-model is constructed based on the plaintext/ciphertext and guessed values of the key. The SCA resistance is measured by computing the minimum number of traces necessary to disclose the unknown key [minimum-number-of-traces-to-disclosure (MTD)]. A higher MTD implies a stronger SCA resistance.

The CEMA-based approaches measure the ability to extract an unknown key by an adversary, and hence, to a certain extent depend on the \textit{adversary's effort} i.e. the number of measurements, the complexity of the attack models and statistical tests used for the attack. From a designer's perspective, it is more crucial to understand whether the measured signatures are correlated to the internal data, irrespective of the outcome of a CEMA attack. We use TVLA as suggested by Goodwill et. al. ~\cite{gilbert2011testing} as a leakage test where the tester selects the key and set of key-specific plaintexts to understand the data-dependency in the captured signatures. We also used higher order statistical moments in t-test to increase the probability of detection, as suggested by Moradi et al. ~\cite{schneider2015leakage}. 

We use a semi-fixed dataset of 100,000 plaintexts for TVLA. A sliding window of 200ns is used for analysis. For CEMA, 500,000 traces were captured for each different configuration of IVR and AES and a sliding window of 80ns is used. A small sliding window for CEMA ensures a better alignment of the filtered traces. The peak correlation is calculated across all filter bands and all-windows to determine the outcome of the attack. %% SM - a very briefl description of exactly how many traces were used, how different attacks were performed etc. Just present the numbers without any analysis as attack methods have already been iniotrduced.

\begin{figure}[!tbhp]
\setlength{\abovecaptionskip}{0pt}
\setlength{\belowcaptionskip}{0pt}
\includegraphics[width=3.3in]{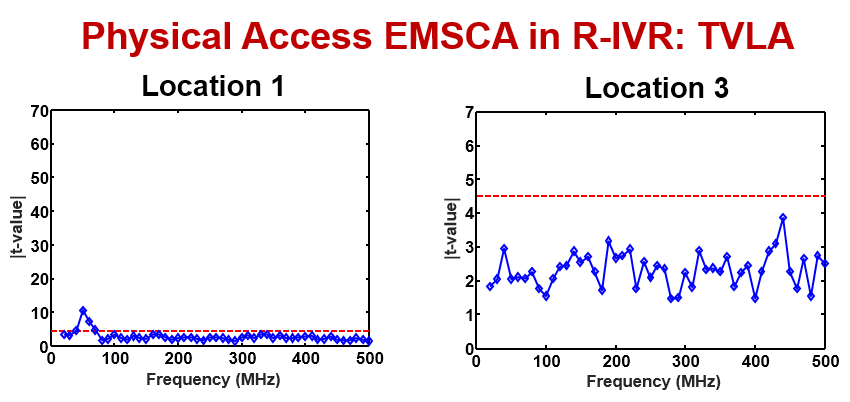}
\caption{TVLA result in R-IVR mode in EMSCA with physical access}
\label{fig:forensics_results_LR}
\end{figure}

\textbf{Signal Post-processing}: A naive adversary aims to mount the SCA on the raw EM signals captured by the probes. However, any misalignment introduced in the chip can thwart such attack and effectively makes an inductive IVR useful in EMSCA protection. However, any skilled adversary would post-process the data before mounting attacks. We assume the role of a skilled adversary and perform necessary post-processing on the captured traces. As the EM signature from a chip with an inductive IVR is a superposition of different sources of EM emission, it is important to properly filter and align the traces before performing any statistical analysis. Another reason why filtering is critical for EMSCA is to extract the useful signature from a coupled EM emission where the EM leakage from the source is coupled with a strong carrier ~\cite{rohatgi2009electromagnetic}. The post-processing step involves filtering and alignment: filtering removes unwanted noise as well as demodulates any modulated signatures and alignment ensures that the same execution step happens across all the captured traces at a given time point. To align the captured traces, we use bandpass filters with bands sliding from 30MHz up to 500MHz in steps of 10MHz. This also replicates the action of a tunable receiver or a demodulator often used in a low-cost EM attack ~\cite{rohatgi2009electromagnetic}. The filtered signals are aligned using cross correlation with the offset limit bounded by the filtering frequency.

\section{EMSCA with Physical Access}  
\label{sec:em-characterization-forensic}

The evaluation of the EMSCA with physical access is performed through TVLA on the traces captured for HP-AES encryptions. An EMSCA with physical access removes any constraints of choosing a probe location and therefore can significantly increase the analysis time, as multiple pins and traces in the package can be probed for EMSCA. As TVLA is generally considered to be a better indicator of leakage compared to CEMA for the same number of measurements, analysis is limited to TVLA on HP-AES. 

{\bf Standalone AES:} 
%\label{off-chip-VRM}
In standalone mode, the high resolution probe is placed in location 1, near the supply (V\textsubscript{DD,AES}) and ground (V\textsubscript{SS,AES}) line. The supply and the ground current of the AES flow through the bondwires marked in pink and blue respectively. The time-domain signatures picked up by the probe in this condition are shown in the Figure \ref{fig:forensics_results_compiled}. The rounds of the HP-AES operation can be clearly identified from the captured waveforms. Figure \ref{fig:forensics_results_compiled} also shows the TVLA results in these conditions against the frequency bands used for filtering. As expected, the t-value crosses the threshold of 4.5 at multiple frequencies, clearly indicating signs of leakage.  

\begin{figure*}[tbhp]
\setlength{\abovecaptionskip}{0pt}
\setlength{\belowcaptionskip}{0pt}
\centering
\includegraphics[width=5.8in]{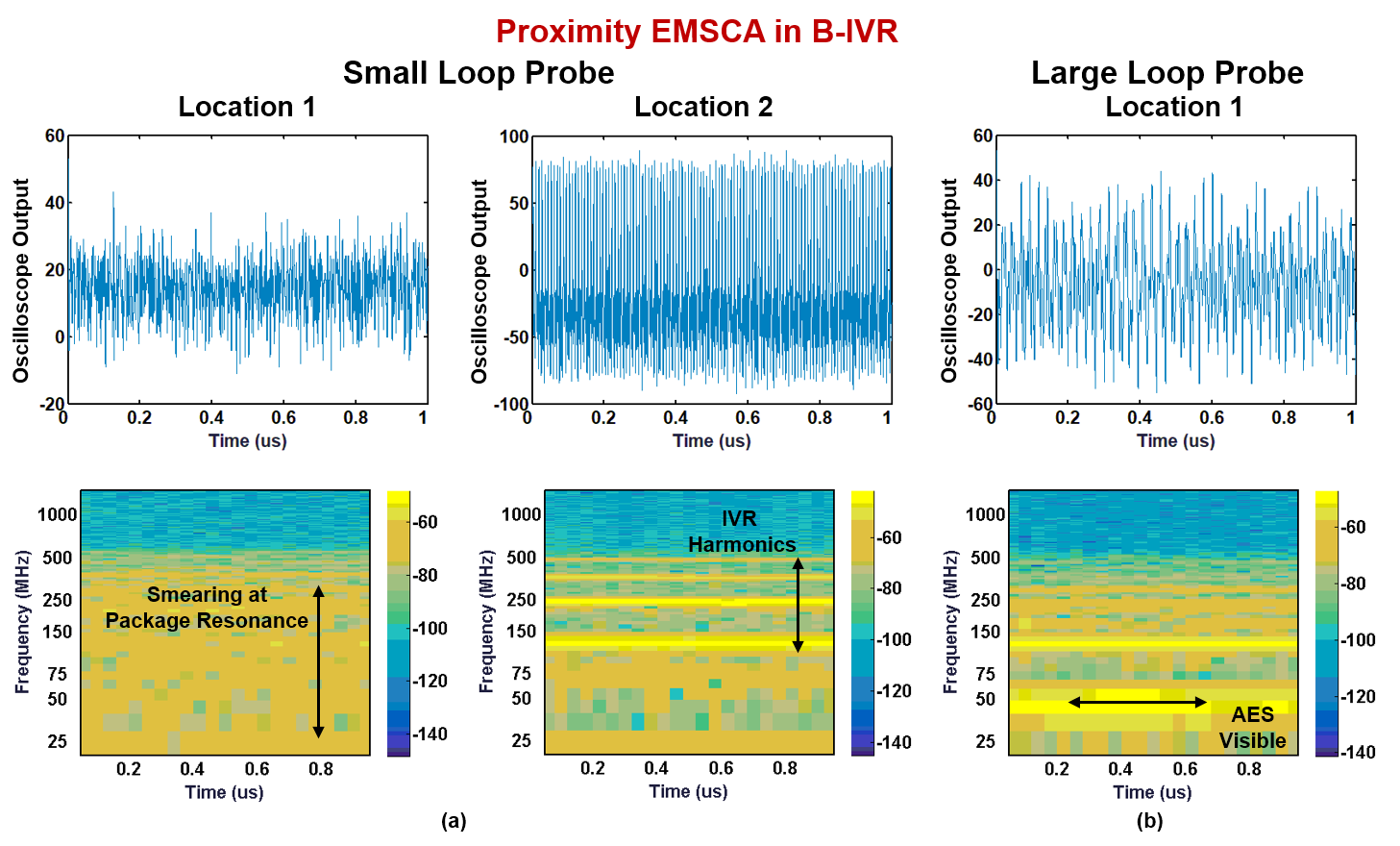}
\caption{EM signatures captured with the passive probes for a HP-AES encryption in the B-IVR mode (a) the small loop probe and (b) the large loop probe}
\vspace{-4mm}
\label{fig:12}
\end{figure*}

{\bf B-IVR:}
%\label{AES-IVR}
When the AES engine is supplied by the IVR, V\textsubscript{DD,AES} is disconnected. But two new locations in the ASIC can potentially emit compromising EM radiation: the IVR input (V\textsubscript{IN,IVR}) and inductor node (V\textsubscript{IND}). The signature picked up near the inductor node (V\textsubscript{IND}) is shown for illustration in Figure \ref{fig:forensics_results_compiled}. The HP-AES operation cannot be visually identified both in time-doamin as well as in spectrogram. However the TVLA results show signs of leakage at both V\textsubscript{IN,IVR} and V\textsubscript{IND} nodes. 

%%SM - The signature picked near V\textsubscript{SS,AES} node is similar for both standalone and IVR-AES modes and shows leakage in both modes.  

%%the AES ground (V\textsubscript{SS,AES}), 

The signature picked up by the high resolution probe is conductive EM emission which is caused due to the current passing through the corresponding pins. Therefore the behavior of the captured signatures from a pin or a trace would be similar to the PSCA properties of the current flowing through the corresponding nodes. This property can explain the observations above. According to the PSCA results shown in ~\cite{kar20178}, the B-IVR input shows TVLA leakage in power signature, which is also observed here. Clearly IVRs can impact the conductive EM emissions in the same way as PSCA behavior. 

{\bf R-IVR:}
The TVLA results on the inductor node (V\textsubscript{IND}) and the supply node (V\textsubscript{DD,IVR}) do not show any leakage in the R-IVR mode. Figure \ref{fig:forensics_results_LR} shows the TVLA data on the inductor node (V\textsubscript{IND}). Although V\textsubscript{SS,AES} won't be accessible for a commercial chip, we performed a TVLA for the purpose of characterization.

\begin{figure*}[!tbhp]
\setlength{\abovecaptionskip}{0pt}
\setlength{\belowcaptionskip}{0pt}
\centering
\includegraphics[width=6.0in,keepaspectratio]{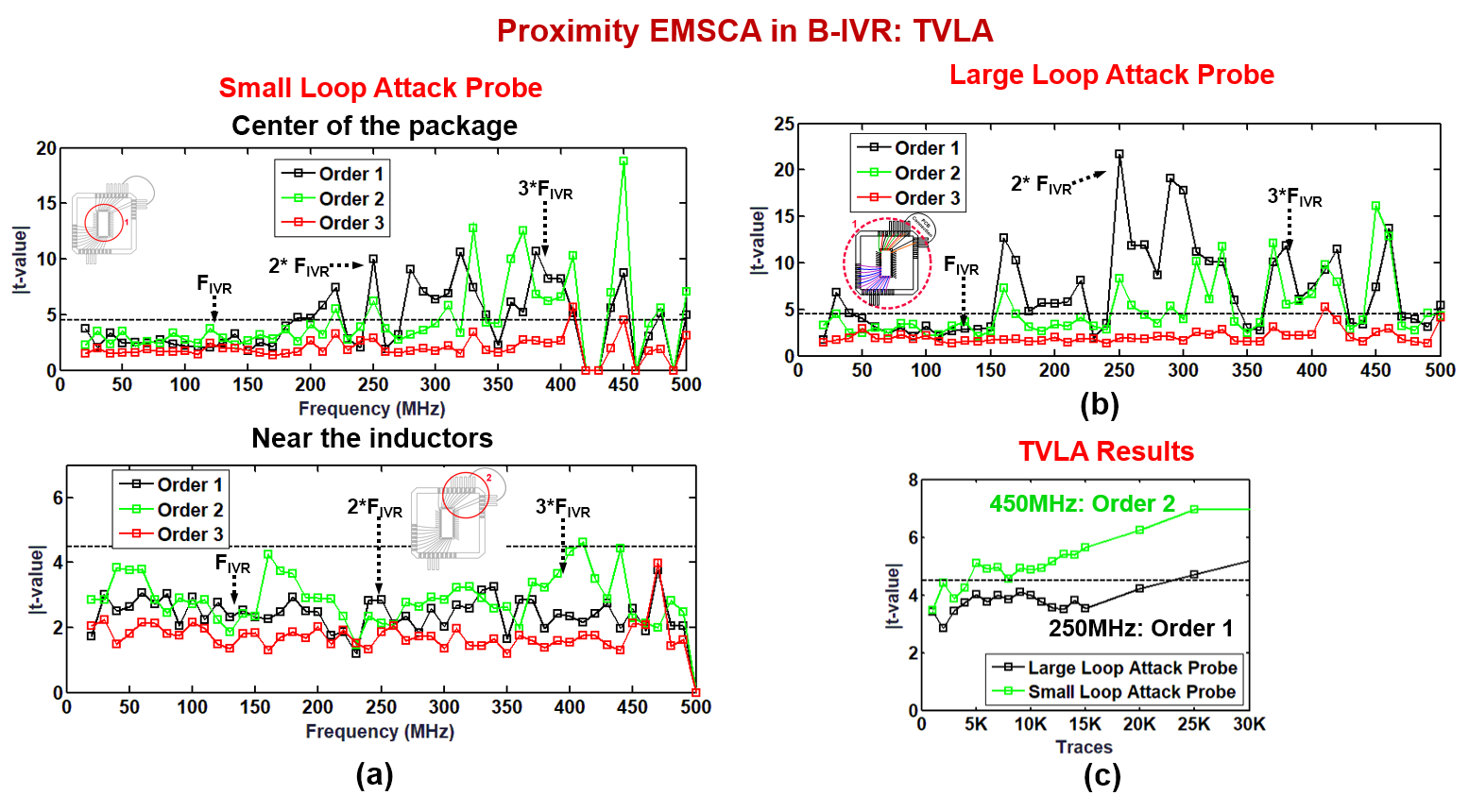}
\caption{Proximity EMSCA using TVLA on HP-AES powered by B-IVR at (a) the locations for the small loop probe and (b) the large loop probe and (c) peak t-value against traces used}
\vspace{-4mm}
\label{fig:15}
\end{figure*}

If a skilled adversary gains physical access to the device, minor design components like sensitive current carrying traces, package pins or supply decoupling capacitors (exploited by authors in ~\cite{balasch2015dpa}) can be exploited for EMSCA. Attacks with physical access have mostly been performed on commercial single-board-computers or microcontrollers ~\cite{longo2015soc, balasch2015dpa}. For the prototype under consideration, we identified the package pins which can potentially emit exploitable EM signatures and measurements using a high resolution probe show that the R-IVR can protect against information leakage in EM signatures. 

One of the security drawbacks of general-purpose products is that the design of the package and the PCB are often agnostic of EM emissions. For example, the normal practice in IVR design is to internally connect the grounds node of AES (V\textsubscript{SS,AES}) and IVR ((V\textsubscript{SS,IVR}), to ensure AES currents flows to ground via the IVR. However, our prototype had the ground node of the AES  (V\textsubscript{SS,AES}) available as an external pin to perform forensics on the chip operation. We have observed that making V\textsubscript{SS,AES} available as a pin is a weak link for Physical Access EMSCA. The signature picked near V\textsubscript{SS,AES} node is similar for both standalone and B-IVR modes and shows leakage in both modes. Even with R-IVR mode, V\textsubscript{SS,AES} does show leakage in the TVLA test (Figure \ref{fig:forensics_results_LR}). This is expected as R-IVR adds minimal noise at the supply node of the AES. Therefore, the AES current flowing through V\textsubscript{SS,AES} remains unchanged. Therefore, our measurement reaffirms that internally connecting the grounds node of the AES engine with that of the IVR, which is a standard practice for commercial products, is necessary to secure the benefits of the R-IVR under Physical Access attack. However, in the next section, we will show that external availability of V\textsubscript{SS,AES} node does not play a major role for Proximity EMSCA.

%\begin{figure*}[!tbhp]
%\includegraphics[width=5in,keepaspectratio]{Figures/Fig14}
%\caption{TVLA (100K traces) with AES powered with the external VRM with (a) the small loop and (b) the large loop probe (c) Peak t-value against traces used}
%\label{fig:14}
%\end{figure*}

\section{Proximity EMSCA on B-IVR}
\label{sec:results}
The proximity EMSCA assumes that the attacker can be in close proximity of the target device, which is the most realistic attack scenario. As HP-AES is more robust to a CEMA, we used TVLA and CEMA for experiments on HP-AES, whereas only CEMA was used for experiments on LP-AES.

\subsection{Characterization of Passive Probes}
\label{probe-characterization}
In standalone mode, the individual rounds of one HP-AES encryption can be captured with both these probes. The signatures picked up by the probes in a B-IVR configuration are shown in Figure \ref{fig:12}. Interestingly, for both locations of the small loop probe, no visible signature of the AES rounds can be identified in the spectrogram and both locations pick up components at package resonance. As the probe is moved from location 1 to location 2, components at the IVR clock frequency and its harmonics increase due to proximity to the inductance. For the large loop probe, the AES operation is visible in the spectrogram. The probe bandwidth of the large loop probe attenuates the IVR clock and its harmonics and significantly increases SNR of the measurement.

\subsection{High Performance AES (HP-AES)}
\subsubsection{TVLA Results}
\label{tvla}
We used a semifixed dataset for TVLA and results using upto 3rd order statistics is computed. A t-value more than 4.5 for an input data-set containing more than 10,000 traces signifies 99.9999\% confidence.

{\bf Standalone HP-AES:}
%\label{tvla-aes-VRM}
We start with the AES engine supplied by the off-chip VRM. Signatures are captured for both the passive probes placed in location 1 which is the middle of the chip. Signatures captured by each of the probes show t-value more than 4.5, clearly showing that the EM signature contains leakage (Figure \ref{fig:15}). The minimum number of traces needed to cross a t-value of 4.5 was 2,000 for both the probes. This experiment clearly shows that the unprotected AES has significant EM information leakage. The component at the IVR frequency is easily filtered out by the post-processing step. Therefore \textit{having a strong EM radiator near the encryption engine isn't effective to protect against EMSCA}.

{\bf B-IVR and HP-AES}
%\label{tvla-aes-IVR-noLR}
We didn't observe any positive TVLA on the raw EM traces without performing any post-processing. After post-processing, the t-values are plotted against the center frequency of the band-pass filters. Although the AES operation cannot be visually distinguished from signatures at location 1 using the small loop probe, TVLA shows leakage at frequencies higher than 200MHz. This is due to the fact that although the low frequency signatures are stronger, they can easily be modulated, whereas the weaker high frequency signals are unmodulated. However, the same probe placed at location 2 i.e closer to the inductance shows weak leakage at higher frequency. The possible reason of this behavior is stronger signature obfuscation by the inductor due to proximity of the probe to the bondwires. Another interesting observation is that the 2nd order TVLA yielded higher t-value than the first order. As the EM signatures are transformed by the nonlinearity of the IVR, higher order statistics can be more effective.
\begin{figure}
\setlength{\abovecaptionskip}{0pt}
\setlength{\belowcaptionskip}{0pt}
\centering
\includegraphics[width=3.3in]{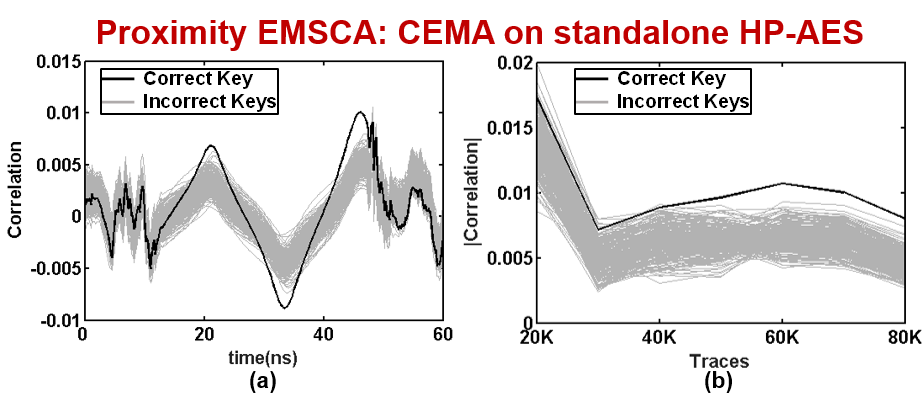}
\caption{CEMA on HP-AES powered with the external VRM for 100,000 traces (a) Correlation against time for byte 10 (b) Correlation vs traces}
\vspace{-4mm}
\label{fig:17}
\end{figure}
For the large loop probe, leakage is observed at the filter band centered at the AES clock frequency as well as the IVR clock frequency and its harmonics. Although the gain of the large loop probe drops significantly after 100MHz as shown in Figure \ref{fig:15}, the larger loop area helps to pick up signatures at higher frequency successfully, leading to TVLA leakage.

We also characterized the minimum number of traces to cross the threshold of 4.5 for each probe at the frequency band and TVLA order which showed highest leakage. The smaller loop needs only ~2,500 traces to cross the 4.5 threshold using a 2nd order TVLA. This is marginally better than the standalone AES and suggests that the obfuscation by the IVR has little effect. One possible reason can also be the placement of the probe away from the inductance. The larger loop requires 20,000 samples to cross the 4.5 threshold. These results are consistent with the observation of authors in ~\cite{kar20178} which found that the B-IVR mode shows leakage in power signatures.

\subsubsection{CEMA Results}
\label{CEMA}
The power-model for CEMA is chosen to be the Hamming distance between the intermediate state at the end of the 9th and the 10th round of the HP-AES. Figure \ref{fig:17} shows the results of CEMA on the HP-AES supplied by the off-chip VRM. A successful CEMA is observed after using 40,000 traces. The corresponding MTD plot is also shown. In B-IVR mode, no successful attack was observed with 500,000 traces. This result matches with the observations in ~\cite{kar20178} where no successful CPA was observed with 100,000 traces at the IVR input.  

\subsection{Low Power AES (LP-AES)}

%\begin{figure}[!tbhp]
%\centering
%\includegraphics[width=3.3in]{./Figures/AES_Serial_Explain_v2.eps}
%\caption{Side channel leakage at IVR input for (a) parallel vs. (b) serial operations of the AES intermediate steps}
%\vspace{-\baselineskip}
%\label{fig:LPAES_explain}
%\end{figure}
\subsubsection{Vulnerability}
We used the Hamming weight of the substitution-box (S-BOX) output in the first round as power-model for attacking LP-AES. SBOX operations on the bytes are executed serially in LP-AES as the engine has only one SBOX hardware. As CEMA targets one byte at a time, the power consumption of rest of the 15 S-BOX operations doesn't appear in the captured signatures unlike HP-AES and therefore the power-model correlates better with the switching activities/EM emission leading to higher vulnerability.

\subsubsection{CEMA}
\begin{figure}[!tb]
\setlength{\abovecaptionskip}{0pt}
\setlength{\belowcaptionskip}{0pt}
\centering
\includegraphics[width=3.3in]{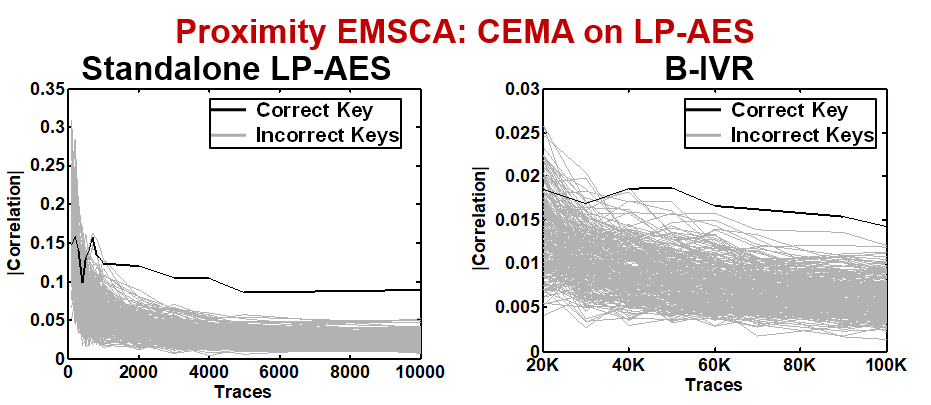}
\caption{CEMA on LP-AES in standalone mode and with B-IVR}
\vspace{-\baselineskip}
\label{fig:LPAES_CEMA}
\end{figure}

\begin{itemize}[leftmargin=*,noitemsep,topsep=0pt]
\item{LP-AES in the standalone configuration is extremely vulnerable to a CEMA as only 1,000 traces are enough to recover a key-byte (Figure \ref{fig:LPAES_CEMA}). This shows that all the sensor nodes, wearables and other edge devices that use a serialized lightweight AES implementation ~\cite{conti2017iot}, are vulnerable to an EMSCA using cheap EM probes.} 
\item{If the B-IVR supplies the AES, the resistance to a CEMA increases by 30x as 30,000 traces were required for successful recovery of a key-byte. A successful CEMA also shows that the system EM signature which is a complex superposition of the EM leakage from the LP-AES and the EM emission from the inductor, is vulnerable against a traditional Hamming-weight based power-model, without the need of a complex power-model or statistical tests.}
\end{itemize}

\section{Proximity EMSCA on R-IVR}
\label{sec:proximity}
The emission from a B-IVR supplying both the AES engines is vulnerable, as demonstrated through TVLA and/or CEMA in the earlier section. Next we enable the R-IVR mode and reevaluate the EMSCA results with both the AES designs. We only used the small loop probe for the following analysis as the captured signatures showed stronger data-dependency compared to the large loop probe.

\subsection{TVLA}
\begin{figure}
\setlength{\abovecaptionskip}{0pt}
\setlength{\belowcaptionskip}{0pt}
\includegraphics[width=3.3in,keepaspectratio]{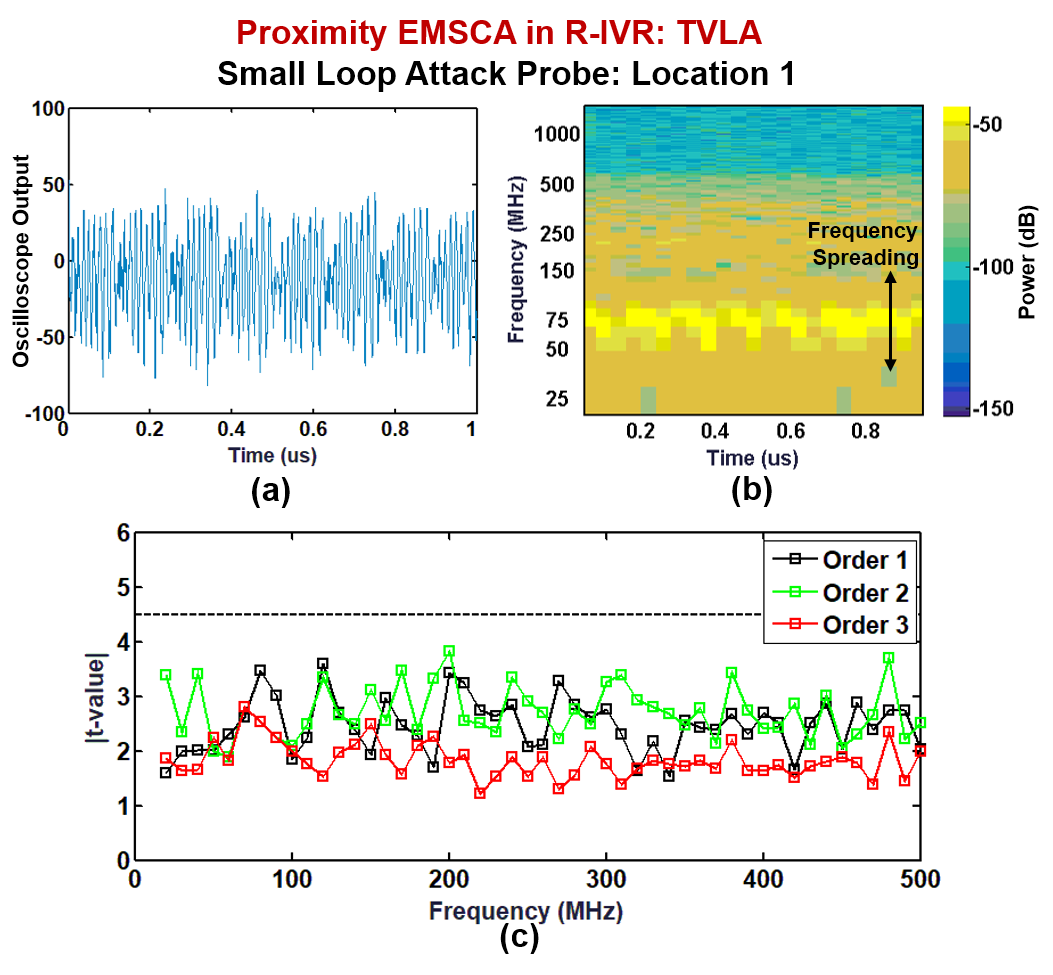}
\caption{Captured signatures in the R-IVR mode in (a) time and (b) spectrogram. (c) TVLA across different frequency bands with 100,000 traces}
\label{fig:16}
\end{figure}
%\label{tvla-aes-IVR-LR}
Figure \ref{fig:16}a and \ref{fig:16}b shows the time domain waveform of the captured signature with small loop probe at location 1 and the corresponding spectrogram when the randomization is enabled in the IVR control loop. As the random delay inserted into the control loop is controlled by a maximal length LFSR, the time domain waveform shows a periodicity dictated by the length of the LFSR. This indirectly creates a frequency spreading or frequency dithering effect with an added degree of randomness. No leakage was observed in the TVLA tests with 100,000 traces (Figure \ref{fig:16}c).

\subsection{CEMA}
\begin{figure}[!tb]
\setlength{\abovecaptionskip}{0pt}
\setlength{\belowcaptionskip}{0pt}
\includegraphics[width=3.3in]{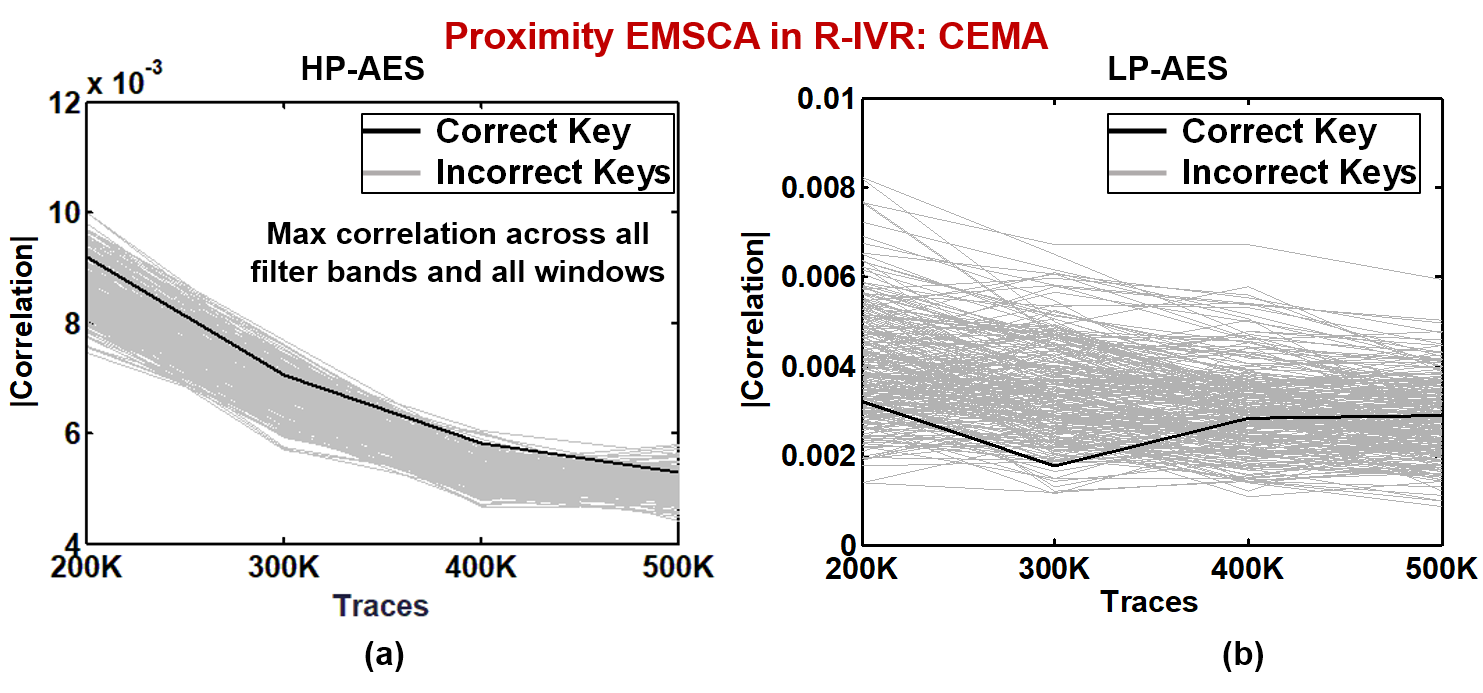}
\caption{CEMA results on the AES designs with R-IVR (a) MTD plot for HP-AES (c) MTD plot for LP-AES}
\label{fig:R_IVR_CEMA}
\end{figure}
\begin{figure*}[!tbhp]
\setlength{\abovecaptionskip}{0pt}
\setlength{\belowcaptionskip}{0pt}
\centering
\includegraphics[width=6in]{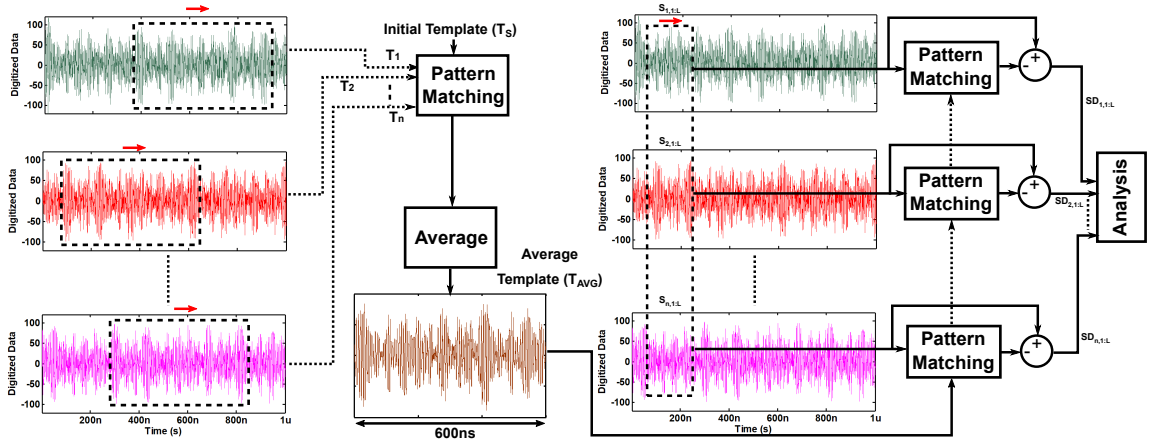}
\caption{Steps of template based CEMA for proximity EMSCA in R-IVR mode}
\vspace{-\baselineskip}
\label{fig:template_attack}
\end{figure*}
CEMA was performed both on HP-AES and LP-AES in the R-IVR mode. No successful attack was observed with 500,000 traces across all bands and all windows (Figure \ref{fig:R_IVR_CEMA}) for both AES designs. HP-AES has a 16X improvement and LP-AES has a 500X improvement in MTD from their respective standalone configurations. 

\subsection{CPA using Templates}
The randomness in the inductor current is manifested in the captured EM signature of the system as shown in Figure \ref{fig:16}. The maximum length 4-bit LFSR which inserts delays proportional to the LFSR output into the IVR control loop, repeats itself after 15 combinations, causing the EM patterns to repeat at a low frequency of $\sim$2MHz. For each measured trace, the LFSR output can be at any one of the 15 possible values and the magnitude of the trace at that point is dependent on that value. Therefore CEMA and TVLA will not be successful unless the effect of the randomization is canceled from the recorded traces. 

We introduce a different attack, referred to as template based CPA, particularly when the LR mode is enabled. The steps of the template based CPA are described in Figure \ref{fig:template_attack}. A template of length 0.6$\mu$s is chosen from a randomly selected trace and patterns of the same length matching with the initial template are found for every trace. All the matched patterns are averaged to generate an average template. We note that the average template contains the EM signature, in absence of any AES operation added with an averaged (over a large number of plaintexts, the leakage at every point will be averaged) EM leakage from the AES operations. Next a window is selected for CEMA which is smaller in length than the template and for each trace, the corresponding portion of the template that matches with the window is subtracted. This generates a set of traces without the steady state variations due to randomization. CEMA is performed on the differential signals, both for the HP-AES and the LP-AES and no successful CEMA was observed with 500,000 traces. 

\section{DISCUSSIONS}
\label{sec:discussion}
\subsection{Robustness Against Attack}
One of the hypothesis for the proposed technique to work is that the EM emission from the AES and the inductor interfere, which is possible when the inductance is integrated closer to the AES engine. With the recent trends in integrated inductor design for power delivery ~\cite{krishnamurthy201720,sturcken2015magnetic}, this seems to be the case. One possible attack mode can be if the adversary access control over the IVR switching frequency. Changing the IVR switching frequency changes the frequency spreading in the R-IVR mode. However typically the switching frequencies of the IVRs cannot be accessed through the firmwares, therefore achieving this requires a destructive and invasive attack. Changing the total load current supplied by the IVR also does not change the switching frequency. Another possible caveat is that EM shielding is added to inductive IVRs to ensure FCC compliance. An EM shielding should attenuate the EM signatures, both from the AES as well as the IVR, and therefore should help in prevention against EM attacks. However tampering the EM shielding might compromise the integrity of the proposed scheme.

\subsection{Public Key Ciphers}
Public key ciphers like ECDH/RSA are widely used for authentication across many devices and have been demonstrated to be vulnerable to EM attacks ~\cite{genkin2016ecdh,genkin2015stealing}. Attacking a public key cipher mainly relies on identifying distinct arithmetic operations like addition/multiplication which is different than SCA on AES where the side channel signatures change over each clock cycle. Therefore attacks on the public key ciphers are typically carried out at much lower frequency bands and can be performed using inexpensive EM probes. The IVR, without and with the randomization scheme, modifies the EM signatures at frequencies $\geq$1MHz, which as demonstrated above is effective for an AES. However as the frequency of interest lies in the KHz range for public key ciphers, the randomization, in its current form, might not be effective. However, one possible solution is to  use a low frequency on-board VRM with a LFSR based control loop randomization, operating at a lower frequency ($\sim$KHz). This will modify the frequency components in the measured EM traces near the frequency of interest and possibly be effective for public key ciphers.

\section{Conclusion}
\label{conclusion}
Protecting EM leakage from modern hardware devices without power and performance penalty and increased packaging cost is a challenging task. Blindside demonstrates that an inductive IVR with a minor design modification can reduce information leakage through EM. The measurement results from the prototype system show that a high-frequency IVR modulates the EM emission from the chip due to presence of an integrated inductance. As an IVR operates at frequencies close to that of a digital processor ($\geq$100MHz), unlike an off-chip VRM module that operate at much lower frequency ($\sim$100KHz), the EM emission from IVR interferes with the EM emission from the AES engines. The system EM signatures, measured using low-cost passive EM probes demonstrate $\geq$13x and 30x improvement in MTD for a high-performance low-latency AES and a low-power low-area compact AES design. If the control loop of the IVR is randomized, $\geq$13x and $\geq$500x improvement in MTD is achieved. As power delivery with integrated inductance is becoming a key component in improving energy efficiency of digital processors, the results show promise in using a common IVR architecture (~\cite{kar20178}) for reducing both power and EM leakage with minimal power, performance, and area overhead.

\end{document}